\documentclass{article}
\usepackage[latin1]{inputenc}
\usepackage{amsmath}
\usepackage{amssymb}
\usepackage{amsthm}
\usepackage{graphicx}

\newcommand{\calg}[1]{{\cal G}_{#1}}
\newcommand{\calm}{{\cal M}}
\newcommand{\pb}{\mu^\mathrm{b}}
\newcommand{\pc}{\mu^\mathrm{c}}
\newcommand{\rswitch}{R_\mathrm{switch}}
\newcommand{\rconv}{R_\mathrm{conv}}
\newcommand{\rw}{R_w}
\newcommand{\rtime}{R_\mathrm{time}}
\newcommand{\bridge}[1]{B^\mathrm{#1}}
\newcommand{\paircut}[2]{C_{(u_{#1} \to u_{#2})}}

\newcommand{\graphDraw}{1.0}
\newcommand{\graphHeight}{3.33cm}

\newtheorem{thm}{Theorem}
\newtheorem{cor}[thm]{Corollary}

\begin{document}

\title{A Study of the Edge-Switching Markov-Chain Method for the Generation of Random Graphs}

\author{Alexandre~O.~Stauffer\\
Valmir~C.~Barbosa\thanks{Corresponding author (valmir@cos.ufrj.br).}\\
\\
Universidade Federal do Rio de Janeiro\\
Programa de Engenharia de Sistemas e Computação, COPPE\\
Caixa Postal 68511\\
21941-972 Rio de Janeiro - RJ, Brazil}

\date{}

\maketitle

\begin{abstract}
We study the problem of generating connected random graphs with no self-loops 
or multiple edges and that, in addition, have a given degree sequence. The 
generation method we focus on is the edge-switching Markov-chain method, whose 
functioning depends on a parameter $w$ related to the method's core operation 
of an edge switch. We analyze two existing heuristics for adjusting $w$ during 
the generation of a graph and show that they result in a Markov chain whose 
stationary distribution is uniform, thus ensuring that generation occurs 
uniformly at random. We also introduce a novel $w$-adjusting heuristic which, 
even though it does not always lead to a Markov chain, is still guaranteed
to converge to the uniform distribution under relatively mild conditions. We 
report on extensive computer experiments comparing the three heuristics' 
performance at generating random graphs whose node degrees are distributed as
power laws.

\end{abstract}

%===============================================================================
%===============================================================================
%===============================================================================
\section{Introduction} \label{sec:intro}

Let $D=\{d_1,d_2,\ldots,d_n\}$ be a set of nonnegative integers such that $d_1 
\geq d_2 \geq \cdots \geq d_n$ and let $\calg{D}$ be the set of all 
connected graphs on $n$ nodes that have no self-loops or multiple edges and for 
which $D$ is the degree sequence. That is, the degree of node $u_j$, $1 \leq j 
\leq n$, is $d_j$. We know from \cite{erdos1960b,berge1989} that $\calg{D}$ is 
a nonempty set, in which case we say that $D$ is realizable, if and only if 
all the following conditions hold: 
\begin{itemize}
   \item $\sum_{j=1}^n d_j$ is even.
   \item $\sum_{j=1}^n d_j \geq 2(n-1)$.
   \item $\sum_{j=1}^k d_j \leq k(k-1) + \sum_{j=k+1}^n \min\{k,d_j\}$  
   for all $k$ such that $1 \leq k \leq n$.
\end{itemize}
We consider in this paper the problem of generating graphs of $\calg{D}$ 
uniformly at random when $D$ is realizable.

In the absence of the connectivity constraint, the problem of generating random 
graphs for a given degree sequence is closely related to some other problems, 
like generating a $(0,1)$ matrix with given marginals \cite{rao1996}, 
approximating the permanent of a matrix \cite{jerrum2004}, and sampling a 
perfect matching or an $f$-factor of a graph \cite{gabow1983,anstee1985}. 
However, it remains generally unknown how to generate graphs uniformly at 
random for a given degree sequence within reasonable time bounds 
\cite{sinclair1993}, even though exceptions exist for some special cases, like 
regular \cite{kim2003} and bipartite graphs \cite{gabow1983,jerrum2004}.

The problem of generating random graphs has recently acquired considerable 
prominence from a practical perspective. Since many real-world networks, like 
the Internet, the WWW, social networks, and scientific-collaboration networks, 
typically have a very large number of nodes and have evolved over time in such 
an unorganized way that only limited information is known about their 
topologies \cite{albert2002,newman2003b}, many studies of their properties have 
been conducted within a random-graph framework 
\cite{adamic2003,gkantsidis2005}. In addition, these networks are now known to 
differ sharply from the classical random-graph model introduced by Erd\H{o}s 
and Rényi \cite{erdos1959,bollobas2002}, in which the node-degree distribution 
is the Poisson distribution. Some empirical studies suggest that many of them 
have node-degree distributions that seem to conform to a power law 
\cite{faloutsos1999,barabasi1999,albert2002,newman2003b}, that is, the 
probability that a randomly chosen node has degree $a$ is proportional to 
$a^{-\tau}$ for some $\tau>1$. 

Clearly, any method for sampling a graph uniformly at random from $\calg{D}$ 
for a given $D$ can be easily extended to generate random graphs having a 
power-law node-degree distribution. We first obtain $D$ by sampling each $d_j$ 
from the power-law distribution. If $D$ turns out not to be realizable, then we 
discard it entirely and obtain a new one, repeating this process while needed. 
We then select the desired graph uniformly at random from $\calg{D}$. 

Other, more complex methods for generating random graphs having node degrees 
distributed as a power law have been proposed. In these methods, generation is 
achieved by successively adding nodes and edges to the graph in such a way that 
tries to follow some principles, like preferential attachment, that are 
believed to have guided the evolution of some real-world networks 
\cite{medina2000,bollobas2003}. However, simply generating a graph having
a given degree sequence sampled from the power-law distribution has been 
observed to perform satisfactorily with regard to certain measures 
\cite{tangmunarunkit2002}. Moreover, this approach can be used to obtain random 
graphs having any node-degree distribution, which is an important flexibility 
since correctly determining the node-degree distribution of real-world 
networks has remained essentially an open problem \cite{achlioptas2005}.

Given a realizable $D$, we consider the generation method that we call the 
edge-switching Markov-chain (ESMC) method for choosing graphs from $\calg{D}$ 
uniformly at random, also variously known by other denominations 
\cite{gkantsidis2003}. This method, which can be modeled as a Markov chain and 
whose details are more thoroughly described in Section~\ref{sec:mc}, employs an 
operation that we call an edge switch to transform a graph of $\calg{D}$ into 
another graph, maybe not in $\calg{D}$ by virtue of not being connected, that 
has the same degree sequence $D$. Let $G$ be the graph being generated. To 
avoid generating unconnected graphs, we periodically perform a connectivity 
test on $G$. If $G$ is unconnected, we undo all the edge switches performed 
since the previous connectivity test. Basically, the method consists of first 
obtaining a graph $G$ from $\calg{D}$ deterministically and then applying a 
series of edge switches and connectivity tests to $G$ until a certain halting 
condition is satisfied. We also discuss in Section~\ref{sec:mc} a methodology 
for obtaining the halting condition, which ultimately also embodies a criterion 
for estimating how close $G$ is to a uniformly random sample from $\calg{D}$.
 
The ESMC method is intrinsically based on an integer parameter $w \geq 1$ giving
the number of edge switches to be attempted between successive connectivity
tests. Naturally, setting $w$ appropriately is crucial to the performance of the
method. When $w$ is too small, a large number of connectivity tests is
performed, which dramatically increases the running time of the method, as the
time complexity of a connectivity test is high in comparison to the time
complexity of an edge switch. On the other hand, when $w$ is excessively large
the probability that the connectivity test is performed on an unconnected graph
tends to be high, possibly causing many edge switches to be undone. Obtaining
an ideal value for $w$ beforehand seems to be an elusive goal, so heuristics
have been proposed for adjusting $w$ along the algorithm's execution
\cite{gkantsidis2003,viger2005}. We discuss the existing heuristics, and also
introduce a new one, in Section~\ref{sec:heur}.
 
We present in Section~\ref{sec:sim} the results of extensive computer
experiments for degree sequences sampled from power-law distributions. We
evaluate the three heuristics described in Section~\ref{sec:heur} along with two
different halting conditions. In general, our computational results indicate
that, on average, our heuristic outperforms the two existing heuristics in terms
of the total running time by a margin of $12\%$ to $86\%$. We conclude in
Section~\ref{sec:conclusion}.

%===============================================================================
%===============================================================================
%===============================================================================
\section{The ESMC method} \label{sec:mc}

\begin{figure*}[!t]
   \centering
   \hspace{\stretch{1}}
   \includegraphics[scale=\graphDraw]{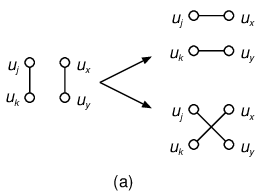} \hspace{\stretch{1}}
   \includegraphics[scale=\graphDraw]{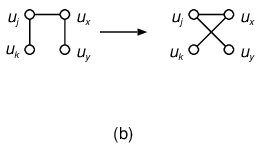} \hspace{\stretch{1}}
   \caption{The two possible edge switches on the edges $(u_j,u_k)$ and 
   $(u_x,u_y)$ (a) and a scenario in which only one of them can be carried 
   through (b).}
   \label{fig:switch}
\end{figure*}

We henceforth denote by $G$ the graph being generated, that is, the graph on 
which the edge switches and the connectivity tests are performed. An edge 
switch is performed on a pair of nonadjacent edges (i.e., edges that share no 
nodes) and consists of removing them from $G$ and adding back one of two other 
pairs of edges. The pair of edges to be added to $G$ is chosen at random from 
these two and the edge switch is only carried through if neither edge of the 
chosen pair already exists in $G$. For example, let $(u_j,u_k)$ and $(u_x,u_y)$ 
be two nonadjacent edges of $G$. The edge-switching operation on $(u_j,u_k)$ 
and $(u_x,u_y)$ consists of removing these edges from $G$ and adding to $G$ 
either $(u_j,u_x)$ and $(u_k,u_y)$ or $(u_j,u_y)$ and $(u_k,u_x)$. Although 
node degrees are clearly seen to remain unchanged by an edge switch, $G$ may 
become an unconnected graph. Figure~\ref{fig:switch}(a) illustrates the two 
possible edges switches on the edges $(u_j,u_k)$ and $(u_x,u_y)$. 
Figure~\ref{fig:switch}(b) illustrates a situation in which only one edge 
switch can be carried through on those edges.

The ESMC method is best described on a Markov chain $\calm$ having one state 
associated with each graph of $\calg{D}$. If 
$G_1,G_2,\ldots,G_{\left|\calg{D}\right|}$ are the graphs in $\calg{D}$ and 
$X_1,X_2,\ldots,X_{\left|\calg{D}\right|}$ are the states of $\calm$, then we 
let, for $1 \leq i \leq \left|\calg{D}\right|$, $X_i$ be the state in which 
$G=G_i$. In essence, the ESMC method consists of initially obtaining a graph of 
$\calg{D}$ and then performing a sequence of transitions on $\calm$ from the 
corresponding state until a certain halting condition is satisfied. 

In order to obtain the initial graph, we employ the Havel-Hakimi algorithm 
\cite{havel1955,hakimi1962,berge1989}, which successively adds edges to an 
initial graph $G$ having $n$ isolated nodes. For $1 \leq j \leq n$, along the 
process let the residual degree $r_j$ of $u_j$ be the difference between $d_j$ 
and the number of edges already incident to $u_j$; clearly, $r_j=d_j$ initially.
The algorithm repeatedly selects the node, say $u_k$, having the highest 
residual degree and connects it to the $r_k$ nodes having the next highest 
residual degrees, which leads to $r_k=0$ and also to smaller values of the other
nodes' residual degrees. The repetition goes on until $r_j=0$ for all $j$ such
that $1 \leq j \leq n$. At this moment, $G$ has degree sequence $D$ but may be
unconnected. Since $D$ is realizable, $G$ must contain a cycle if it is not
connected. If we take an edge of this cycle and an edge of another connected
component, and perform an edge switch on them, then necessarily two of the
connected components of $G$ are merged together into a single one. This process
can be repeated until $G$ becomes connected. 

Let us then describe what constitutes a transition in $\calm$. Let $w \geq 1$ 
be an integer parameter. A transition in $\calm$ is a sequence of $w$ steps 
that we call edge-switching attempts. In each edge-switching attempt, we 
randomly select two distinct edges of $G$. If they are not adjacent, then we 
randomly choose one of the two possible edge switches. If the chosen edge switch
is feasible, that is, it does not involve adding an edge that already exists in
$G$, then we go on and perform the edge switch. $G$ is kept unchanged otherwise.
After $w$ edge-switching attempts, we perform a connectivity test on $G$. If $G$
turns out to be unconnected, then we undo all the edge switches performed during
the previous $w$ edge-switching attempts. 

Now let $m$ be the number of edges of $G$. If we use an array with the edges of 
the graph, an adjacency matrix, and an appropriate collection of incidence lists
and pointers, then an edge switch can be done in $O(1)$ time while requiring 
$\Theta(n^2)$ space, which for large $n$ is prohibitive. An alternative way is 
to use an array with the edges of the graph and an appropriate  collection of 
incidence trees and pointers, which leads to $O(\log d_1)$ time and $\Theta(m)$ 
space instead. This has been our choice in all the computer experiments we
discuss in Section~\ref{sec:sim}. In any case, and considering that the
connectivity test can be performed in $O(m)$ time, setting $w$ properly is
essential to the achievement of good performance. We return to this issue in
Section~\ref{sec:heur}.

In $\calm$, a transition exists from $X_i$ to $X_j$, $1 \leq i,j \leq 
\left|\calg{D}\right|$, if and only if there is a sequence of $w$ 
edge-switching attempts transforming $G_i$ into $G_j$. Let $p_{i,j}$ be the 
probability associated with this transition. Clearly, $p_{i,j}=p_{j,i}$ so long 
as $w$ is constant (every edge switch can be undone with the same probability
with which it was previously done), and $p_{i,i}>0$ (every edge-switching
attempt may select adjacent edges to switch or an infeasible edge switch). The
main results that pertain to the use of $\calm$ in sampling a random graph from
$\calg{D}$ uniformly at random are consequences of the following two classic
theorems on Markov chains \cite{karlin1975,mitzenmacher2005}.

\begin{thm}\label{thm:stationary}
   A finite, irreducible, and aperiodic Markov chain converges to a unique 
   stationary distribution regardless of the initial state. 
\end{thm}

\begin{thm}\label{thm:rev}
   Given a finite, irreducible, and aperiodic Markov chain with state space
   $\{Y_1,Y_2,\ldots,Y_k\}$, let $q_{i,j}$ be the probability associated with 
   the transition from $Y_i$ to $Y_j$. If there are nonnegative numbers 
   $\pi_1,\pi_2,\ldots,\pi_k$ such that $\sum_{i=1}^k \pi_i = 1$, and 
   furthermore 
   $$
      \pi_i q_{i,j} = \pi_j q_{j,i}
   $$
   for all $i,j$ such that $1 \leq i,j \leq k$, then the stationary 
   distribution of this Markov chain is given by $\pi_1,\pi_2,\ldots,\pi_k$, 
   with the probability associated with $Y_i$ being $\pi_i$, $1 \leq i \leq k$.
\end{thm}

Corollary~\ref{cor:unif}, given next, follows directly from 
Theorem~\ref{thm:rev}.

\begin{cor}\label{cor:unif}
   If $q_{i,j}=q_{j,i}$ for all $i,j$ such that $1 \leq i,j \leq k$, then the 
   stationary distribution of the Markov chain of Theorem~\ref{thm:rev} is the 
   uniform distribution.
\end{cor}

Our chain $\calm$ is certainly finite and is also irreducible (since there is a 
sequence of transitions between any two states of $\calm$ \cite{taylor1981}) 
and aperiodic (since $p_{i,i}>0$ for all $i$ such that $1 \leq i \leq 
\left|\calg{D}\right|$). Also, $p_{i,j}=p_{j,i}$ for all $i,j$ such that $1 
\leq i,j \leq \left|\calg{D}\right|$ if $w$ is constant. By 
Corollary~\ref{cor:unif}, we then have the following.

\begin{cor}\label{cor:wunif}
   If $w$ is constant, then $\calm$ converges to the uniform distribution 
   regardless of the initial state. 
\end{cor}

We finalize the section by discussing a halting condition for the ESMC method. 
For $t \geq 1$, let $g(t)$ be a function of $G$ right after the $t$th 
transition. Let also 
\begin{equation}
   \bar g(t) = \frac{g(0)+g(1)+\cdots+g(t)}{t+1},
   \label{eq:est}
\end{equation}
where $g(0)$ refers to the initial $G$. The quantity in (\ref{eq:est}) is known 
to give an unbiased estimator of the expected value of $g(t)$ under the 
stationary distribution whenever Theorem~\ref{thm:stationary} holds 
\cite{karlin1975}. We use $\bar g(t)$ as an indirect 
indicator of the convergence of $\calm$. Let $\delta \geq 1$ and $\gamma > 0$ 
be two parameters, the former an integer. Our halting condition after the $t$th 
transition is that the inequality
\begin{equation}
   \left| \frac{\bar g(z) - \bar g(t-\delta)}{\bar g(t-\delta)} \right| \leq \gamma
   \label{eq:haltcond}
\end{equation}
hold right after each of the $\delta$ most recent transitions that precede (with
inclusion) the $t$th one (that is, for $t-\delta+1 \leq z \leq t$). The efficacy
of this halting condition depends clearly on the function $g(t)$. In
Section~\ref{sec:sim} we present computational results for two different choices
of $g(t)$ (deciding to halt based on one of them, however, bears no direct
relationship to deciding to halt on the other).

%===============================================================================
%===============================================================================
%===============================================================================
\section{Heuristics for parameter adjustment} \label{sec:heur}

As we remarked at the end of Section~\ref{sec:intro}, adjusting $w$ along the 
evolution of $\calm$ is a viable alternative, aiming at better convergence 
properties, to fixing its value at the onset. In this section we discuss some 
heuristics to do this. Each transition consists now of performing $w$ 
edge-switching attempts, a connectivity test (with the ensuing possible undoing 
of all the edge switches performed during the $w$ attempts), and moreover an 
update of the value of $w$. We consider two approaches to adjusting $w$. The 
first consists of a mechanism that is used in all existing heuristics for 
adjusting $w$ in accordance with the result of the previous connectivity test. 
The other one is a new heuristic that adjusts $w$ aiming at approximating a 
given probability for the success of the next connectivity test. Notice that, 
in either case, Corollary~\ref{cor:wunif} is no longer applicable and the 
convergence of $\calm$ has to be re-examined.

%===============================================================================
\subsection{Two current heuristics} \label{sec:oldh}

Let us begin with the first approach. We start with $w=1$ and increase the 
value of $w$ whenever the connectivity test succeeds; we decrease it otherwise. 
As we demonstrate next, a Markov chain exists associated with this approach 
that has a uniform stationary distribution.

Let $\calm'$ be a Markov chain whose states are each associated with a graph of 
$\calg{D}$ and a value of $w$. We denote by $X'_{i,a}$ the state of $\calm'$ 
associated with $G_i$ and $w=a$. While $\calm'$ models the approach in question
faithfully, it has more than one state associated with each graph of $\calg{D}$
and using it directly in our analysis may prove cumbersome. We then introduce
another Markov chain, denoted by $\calm''$ and having only 
$\left|\calg{D}\right|$ states, each associated with a graph of $\calg{D}$. We 
denote by $X''_i$ the state of $\calm''$ associated with $G_i$. This state is 
the union of $X'_{i,a}$ for all $a \geq 1$, i.e., $X''_i$ results from 
clustering together all the states of $\calm'$ that correspond to $G_i$.

In order to make the state space of $\calm'$ finite, we limit the value of $w$ 
by a fixed upper bound, henceforth denoted by $W$. This strategy not only makes 
$\calm'$ a finite Markov chain, which is crucial to the analysis that follows, 
but also avoids excessively large $w$ values, which may jeopardize the 
approach's efficacy, especially in relation to the halting condition, as $g(t)$
may end up being calculated too sporadically with respect to the edge-switching
attempts.

It is a consequence of our discussion of Section~\ref{sec:mc} that, in 
$\calm'$, any state $X'_{i,a}$ is reachable from any state $X'_{j,a}$ without 
even going through states for which $w \neq a$. As any of the involved 
transitions corresponds unequivocally to a transition in $\calm''$, it follows 
immediately that $\calm''$ is irreducible and aperiodic. By 
Theorem~\ref{thm:stationary}, $\calm''$ converges to a unique stationary 
distribution.

Now let $X'_{i,a}$ and $X'_{j,b}$ be any two states of $\calm'$. The existence 
of a transition from $X'_{i,a}$ to $X'_{j,b}$ means that there is a sequence of 
$a$ edge-switching attempts transforming $G_i$ into $G_j$ and updating $w$ to 
$b$. Since every edge switch can be undone (as before, with the same probability
with which it was previously done), there is also a sequence of $a$ 
edge-switching attempts transforming $G_j$ into $G_i$ and updating $w$ to $b$
(i.e., from $X'_{j,a}$ to $X'_{i,b}$). If $p''_{i,j}$ is the probability
associated with the transition from $X''_i$ to $X''_j$ in $\calm''$, then
clearly $p''_{i,j}=p''_{j,i}$ and we have the following consequence of
Corollary~\ref{cor:unif}.

\begin{cor}\label{cor:hunif}
   $\calm''$ converges to the uniform distribution regardless of the 
   initial state. 
\end{cor}

Two heuristics for adjusting $w$ based on the outcome of the connectivity test 
have been proposed. In the first heuristic, which is a variation of the one
introduced by Gkantsidis, Mihail, and Zegura in \cite{gkantsidis2003} and is
henceforth referred to as the GMZ heuristic, $w$ is updated to $w+1$ when the
connectivity test is successful and to $\lceil w/2 \rceil$
otherwise.\footnote{The original heuristic in \cite{gkantsidis2003} differs from
this variation in two ways. First, it forces the probability of remaining at the
same state after a transition to be at least $0.5$; secondly, the choice of the
two edges to undergo a switch is restricted to nonadjacent edge pairs only.
However, by adopting our variation of the heuristic, which lets adjacent edge
pairs be chosen as well, the probability of remaining at the same state is
automatically reinforced.} The other heuristic, due to Viger and Latapy 
\cite{viger2005} and henceforth referred to as the VL heuristic, is based on 
two parameters, $q^+$ and $q^-$, such that $q^+ > 0$ and $0 < q^- < 1$. It 
prescribes that $w$ be updated to $(1+q^+)w$ when the connectivity test 
succeeds and to $(1-q^-)w$ otherwise. In \cite{viger2005} it is suggested that 
these two parameters be adjusted in such a way as to satisfy $q^+/q^-=e-1$. We 
report on computer experiments with these two heuristics in
Section~\ref{sec:sim}.

%===============================================================================
\subsection{A new heuristic} \label{sec:newh}

Let $\alpha$ be such that $0 < \alpha < 1$. We introduce a new heuristic to 
adjust $w$ whose goal is to achieve a constant probability $\alpha$ for the 
success of the next connectivity test. The new heuristic relies on a special 
connectivity test, whose details are described in Appendix~\ref{sec:test}, that 
not only checks whether $G$ is connected but also calculates the probability 
that $G$ remains connected after an edge-switching attempt. We refer to this 
new heuristic as the SB heuristic.

For $1 \leq i \leq \left|\calg{D}\right|$, let $\rho_i$ be the probability that 
$G_i$ remains connected after an edge-switching attempt. The SB heuristic is 
based on the assumption that the probability that a connectivity test succeeds 
after $w$ consecutive edge-switching attempts starting at $G_i$ is 
$(\rho_i)^w$. In other words, we assume that the probability that a graph 
remains connected after each of the $w$ edge-switching attempts is $\rho_i$, 
and also that it suffices that one single edge switch yields an unconnected 
graph in order for the next connectivity test to be unsuccessful. We note that 
the latter assumption makes special sense under power-law node-degree 
distributions, since in such cases random node deletions are not likely to split
the graph into more than one relatively large connected component 
\cite{albert2000,cohen2000}. What this means is that, when an edge switch
renders the graph unconnected, the forthcoming connectivity test can only
succeed if a subsequent edge switch is performed on edges from different
connected components, that is, most likely on at least one edge belonging to a
relatively small connected component, which is a low-probability event.

\begin{figure*}[!t]
   \centering
   \includegraphics[scale=\graphDraw]{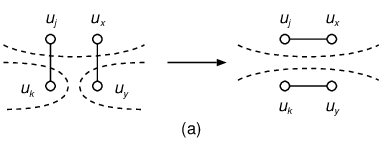} \\
   \includegraphics[scale=\graphDraw]{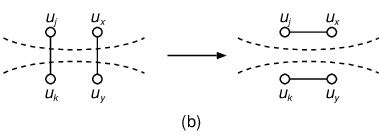} 
   \caption{The two possibilities for an edge switch to produce an unconnected
   graph. In part~(a), $(u_j,u_k)$ and $(u_x,u_y)$ constitute a nonadjacent,
   non-neighbor bridge pair. In part~(b), $(u_j,u_k)$ and $(u_x,u_y)$ constitute
   a nonadjacent, non-neighbor pair cut. The dashed lines delimit the connected 
   components that appear when the edges crossing them are removed from the    
   graph.}
   \label{fig:pbpc}
\end{figure*}

In order to obtain $\rho_i$, we calculate the number of pairs of edges of $G_i$ 
on which performing an edge switch generates an unconnected graph. Let 
$(u_j,u_k)$ and $(u_x,u_y)$ be two edges of $G_i$. We say that $(u_j,u_k)$ and 
$(u_x,u_y)$ are neighbors if at least one other edge joins two of the four 
nodes in $G_i$. Clearly, an edge switch can only make $G_i$ unconnected if the 
two edges involved in the switch constitute a cut of $G_i$. In addition, it is 
also necessary that the edge switch be performed on two edges that are not 
neighbors. Given two nonadjacent edges $(u_j,u_k)$ and $(u_x,u_y)$ that
constitute a cut of $G_i$ and moreover are not neighbors, only one of the two
possible edge switches generates an unconnected graph. This is illustrated in 
Figure~\ref{fig:pbpc}: in part~(a), each edge is, individually, a cut of the 
graph, constituting what we call a nonadjacent, non-neighbor bridge pair; in
part~(b), only together are the two edges a cut of the graph, constituting what
we call a nonadjacent, non-neighbor pair cut.

Clearly, there are $\binom{m}{2}=m(m-1)/2$ pairs of distinct edges, and on each
one we may perform up to two edge switches, depending on how many are feasible.
Let $\pb_i$ be the ratio of the number of nonadjacent, non-neighbor bridge pairs
in $G_i$ to $m(m-1)$. Note that $\pb_i$ gives the probability that we choose a
nonadjacent, non-neighbor bridge pair and perform on it the edge switch that
produces an unconnected graph. Likewise, let $\pc_i$ be the ratio of the number
of nonadjacent, non-neighbor pair cuts in $G_i$ to $m(m-1)$. Then $\pc_i$ is the
probability that we choose a nonadjacent, non-neighbor pair cut and perform on
it the edge switch that produces an unconnected graph. We clearly have
\begin{equation}
   \rho_i=1-\pb_i-\pc_i. 
   \label{eq:rho}
\end{equation}
In Appendix~\ref{sec:test} we give a connectivity test that calculates the 
value of $\rho_i$ and is asymptotically no harder than depth-first search in the
worst case.

If $G_i$ is the graph obtained right after a connectivity test, then the
intuition behind the SB heuristic indicates that $w$ should be adjusted in a way
that led to $\alpha=(\rho_i)^w$, yielding
\begin{equation}
   w = \frac{\ln \alpha}{\ln \rho_i}. 
   \label{eq:w1}
\end{equation}
Notice, however, that each graph $G_i$ of $\calg{D}$ may have a different 
$\rho_i$, so the Markov chain modeling this method might converge to a
stationary distribution that is different from the uniform distribution. For
this reason, we define $\bar \rho(t)$ to be the average of every $\rho_i$
obtained right after each of the first $t+1$ connectivity tests (the initial one
and the $t$ others that correspond to transitions). We then let the SB
heuristic adjust $w$ according to
\begin{equation}
   w = \left\lceil\frac{\ln \alpha}{\ln \bar \rho(t)}\right\rceil
   \label{eq:w}
\end{equation}
right after the $t$th connectivity test. Note, in connection with (\ref{eq:w}), 
that $w$ is assuredly a positive integer. Furthermore, for the reasons 
discussed in Section~\ref{sec:oldh}, we limit $w$ by a fixed upper bound $W$. 

We remark, finally, that as a consequence of $w$ being adjusted as a function of
every $\rho_i$ ever obtained, the method cannot be modeled as a Markov chain
and, to be rigorous, can no longer even be treated as a variation of the ESMC
method in which another heuristic is used. However, if $\bar \rho(t)$
converges as $t \to \infty$, then $w$ also converges. In this case, $w$
approaches a constant and, as noted in Section~\ref{sec:mc}, we once again have
a method that can be modeled as a Markov chain having a uniform stationary
distribution. In Section~\ref{sec:sim}, our approach to assessing the
convergence of $\bar \rho(t)$ (and of $w$, consequently) is to compare the
average value of $g(t)$ at the end of an execution under the SB heuristic to
those obtained under the GMZ and VL heuristics. As we demonstrate in that
section, the figures for the SB heuristic vary within relatively small
percentages with respect to those of either of the other two heuristics and we
take this as indication that $\bar \rho(t)$ is close to convergence. In
what follows, then, we continue to refer to the SB heuristic as an alternative
for use with the ESMC method.

%===============================================================================
%===============================================================================
%===============================================================================
\section{Computational results} \label{sec:sim}

In this section we present computational results for the three heuristics of
Section~\ref{sec:heur}. We have concentrated on power laws with
$\tau=2.0,2.1,\ldots,3.0$ and set $n=10^3$. All experiments were carried out on
a Pentium 4HT running at 3GHz with 1GB of main memory. All running times we
report refer to total elapsed times under a Linux operating system hosting one
single user.

Before discussing our experiments, we pause momentarily to elaborate on a 
curious behavior of the power-law distribution. From Section~\ref{sec:intro}, 
we know that, in order for $D$ to be realizable, its average node degree must be
no less than the average node degree of a tree, which is approximately $2$ for 
sufficiently large $n$. For $n=10^3$, this is expected to hold only for 
$\tau \lessapprox 2.47$, meaning that for $\tau \gtrapprox 2.47$ $D$ is 
expected not to be realizable. By requiring realizability as we repeatedly 
sample $D$ from the power law, we are in fact making the node-degree 
distribution be slightly different from that very power law. What we have 
observed is that, for $\tau \gtrapprox 2.47$, the node-degree variance for 
realizable degree sequences tends to increase with $\tau$ while the number of 
edges remains roughly constant. These characteristics have affected the results 
we present next very strongly.

In our experiments, we used $W=10^4$. For each value of $\tau$, we sampled $600$
realizable degree sequences and, for each of them, executed the generation
method using the three heuristics and two distinct halting conditions. We
carried out the VL heuristic for $q^+=0.1,0.2,0.3$ and set $q^-$ in such a way
that $q^+/q^-=e-1$. The SB heuristic was carried out for $\alpha=0.1,0.2,0.3$.

We have focused on analyzing four indicators, each calculated from the $600$
executions with each heuristic and each halting condition. The first one, which
we denote by $\rconv$, is the ratio of the average $g(t)$ value at the end of
an execution to the average value of $\bar g(t)$ also at the end of an
execution. $\rconv$ can be used as a source of information on the convergence of
the Markov chain, as we know that $\bar g(t)$ is an unbiased estimator for
$g(t)$. Generally, the deviation of $\rconv$ from $1$ grows with how far the
generated graph is from a uniformly random sample of $\calg{D}$. The second
indicator, which we denote by $\rswitch$, is the average number of edge switches
performed during an execution that are not undone as a result of the
connectivity test. The third indicator, which we denote by $\rw$, is the average
value of $w$ at the end of an execution. The last indicator, finally, is the
average running time (in minutes) of an execution and is denoted by $\rtime$.

%===============================================================================
\subsection{Halting on the clustering coefficient}

For the first halting condition, we have let $g(t)$ be the clustering 
coefficient of $G$. This coefficient is the ratio of three times the number of 
triangles in $G$ to the number of three-edge paths in $G$ (each triangle 
corresponds to three such paths) \cite{newman2003b}. Calculating the clustering 
coefficient requires $O(d_1 m)$ time, as a triangle is identified by checking 
whether an edge's end nodes have a common neighbor. We have used $\delta=60$ and
$\gamma=10^{-4}$ for this halting condition.

With regard to our discussion at the end of Section~\ref{sec:newh} on the
convergence of $\bar \rho(t)$, we have observed the average value of $g(t)$
under the SB heuristic to vary within only roughly $5\%$ of the values obtained
for the GMZ heuristic for most values of $\tau$, the exceptions being $\tau=2.4$
($7.8\%$) and $\tau=2.5$ ($6.3\%$). As for the VL heuristic, the percentage
drops to roughly $3\%$, the exceptions being the same with $5.7\%$ for
$\tau=2.4$ and $5\%$ for $\tau=2.5$.

\begin{figure*}[!p]
   \centering
   \begin{tabular}{rrr}
   \includegraphics[height=\graphHeight]{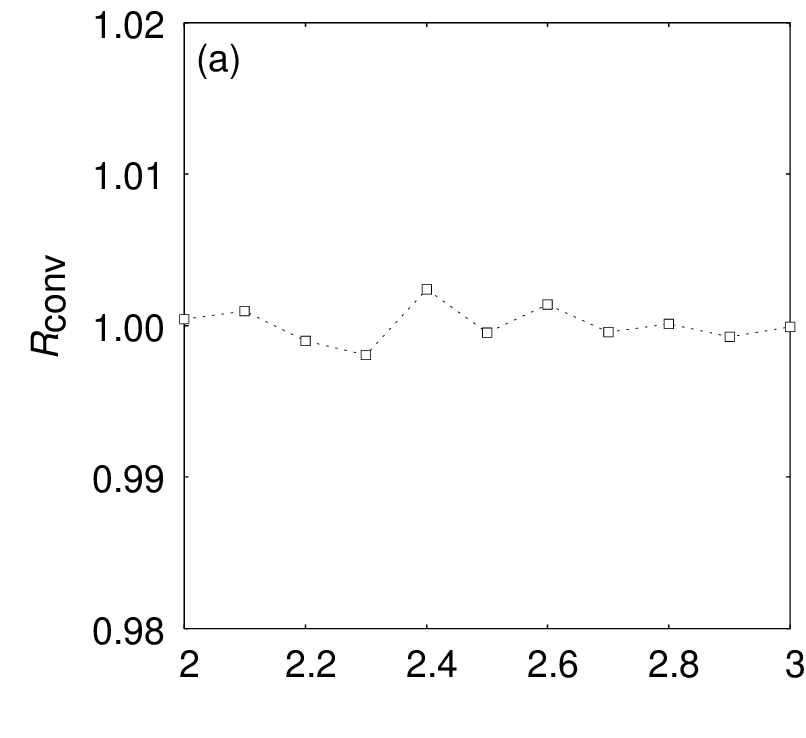} &
   \includegraphics[height=\graphHeight]{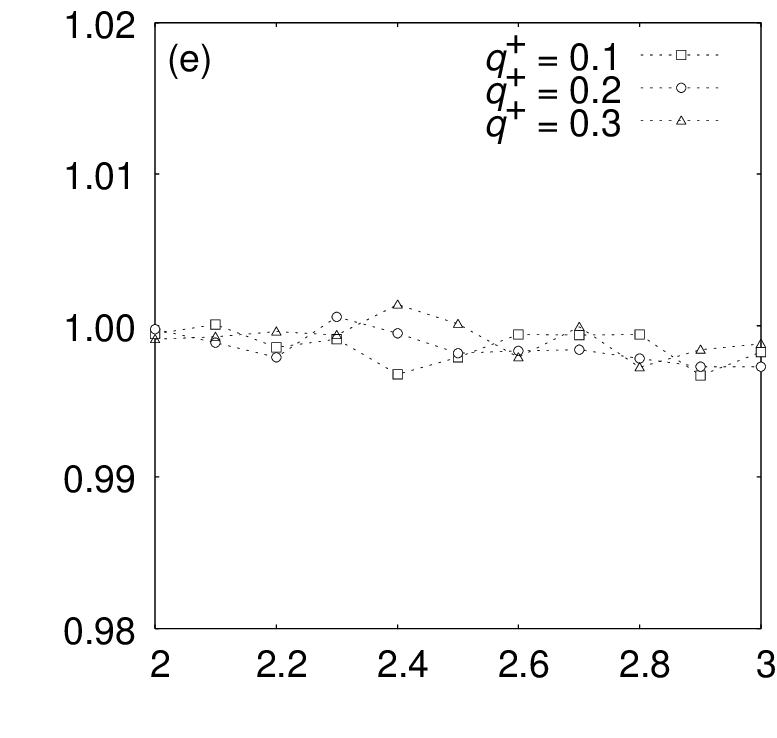} &
   \includegraphics[height=\graphHeight]{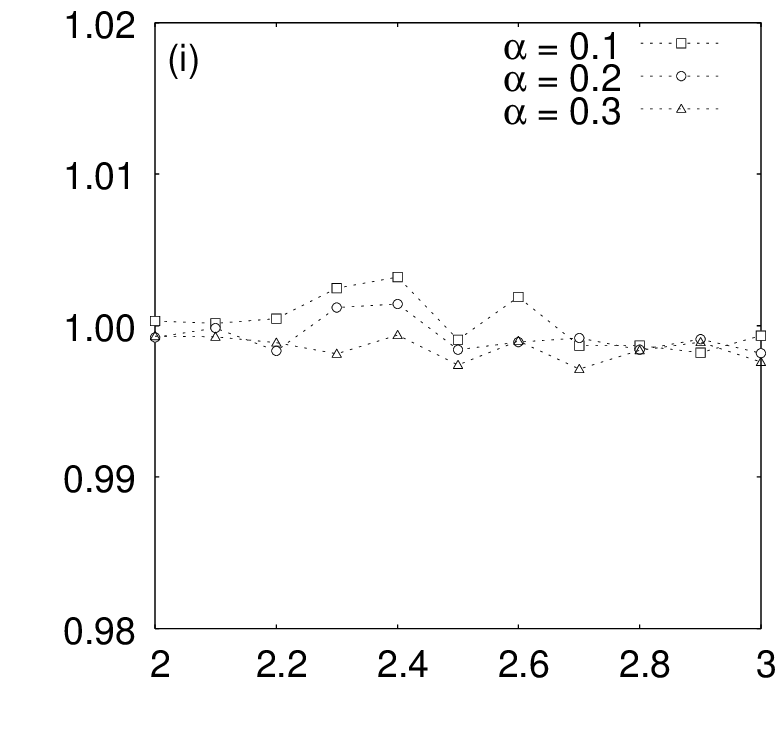} \\
   \includegraphics[height=\graphHeight]{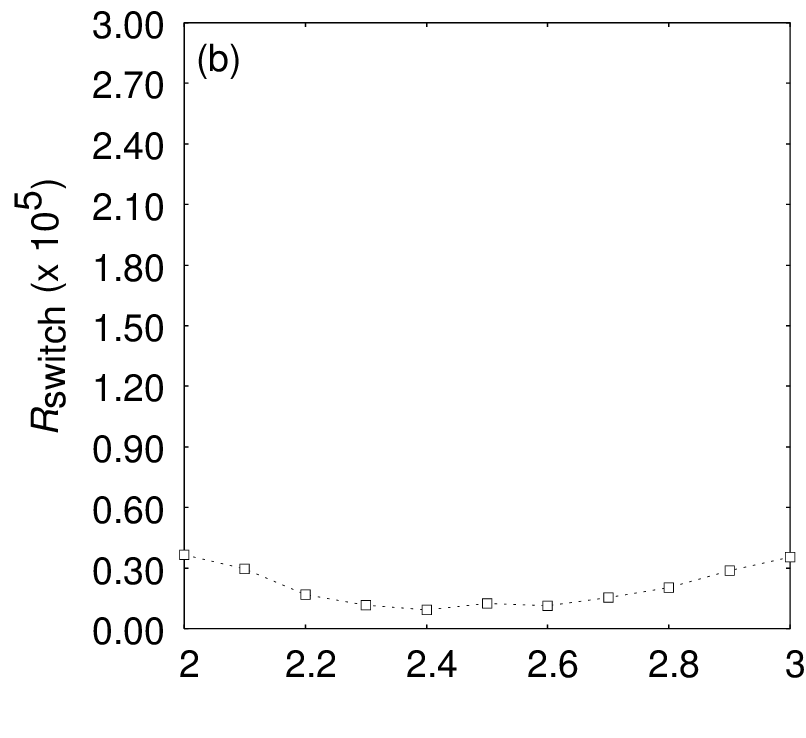} &
   \includegraphics[height=\graphHeight]{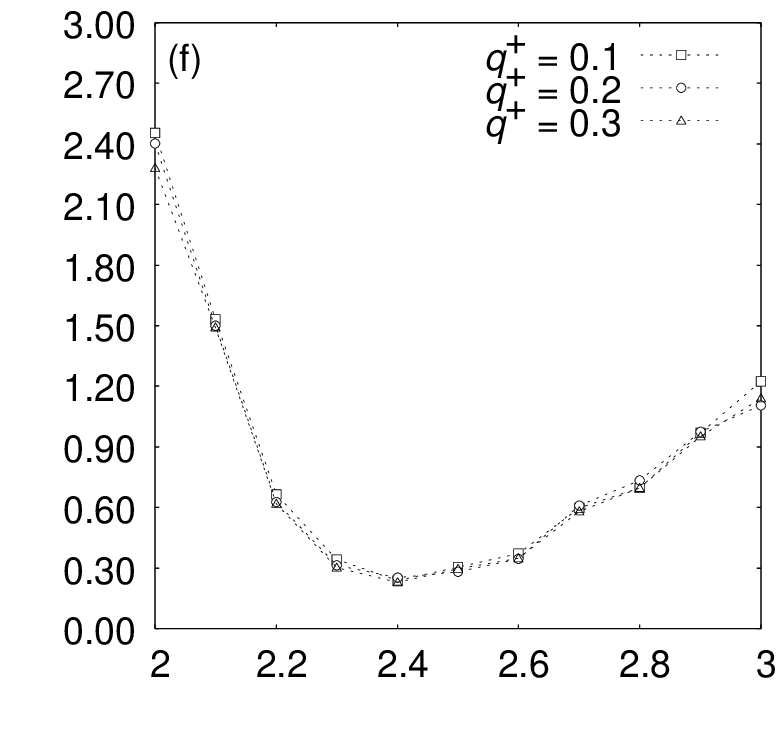} &
   \includegraphics[height=\graphHeight]{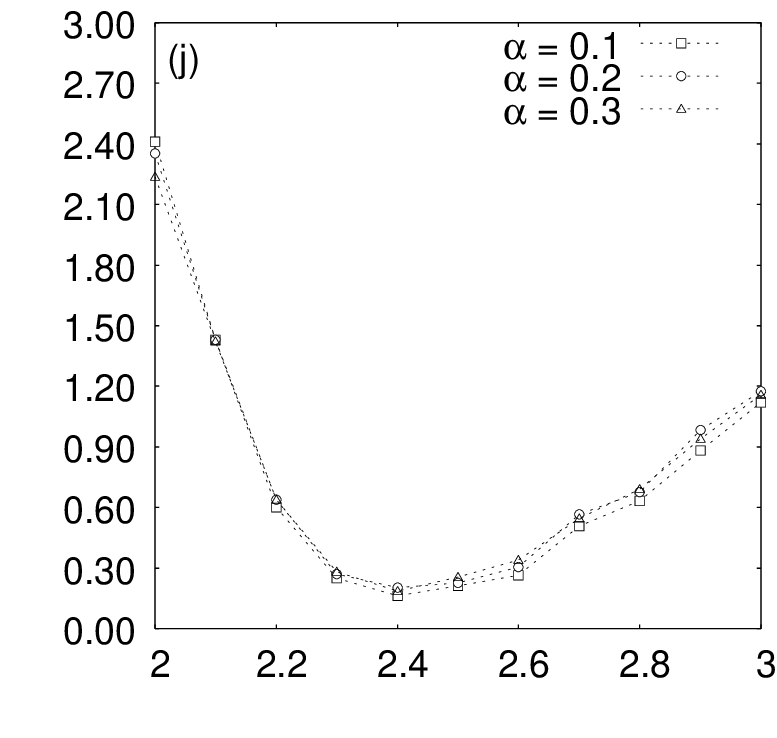} \\
   \includegraphics[height=\graphHeight]{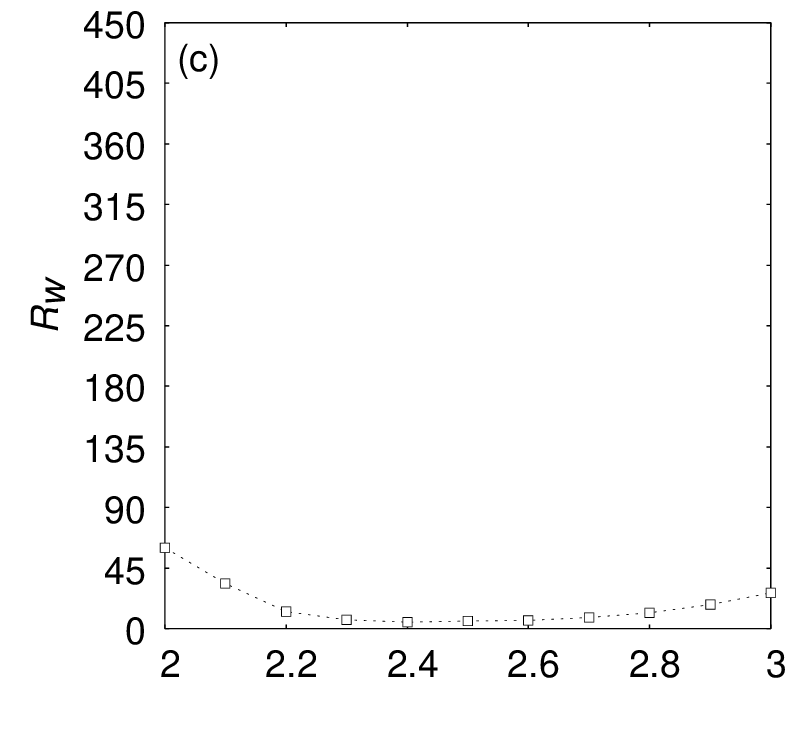} &
   \includegraphics[height=\graphHeight]{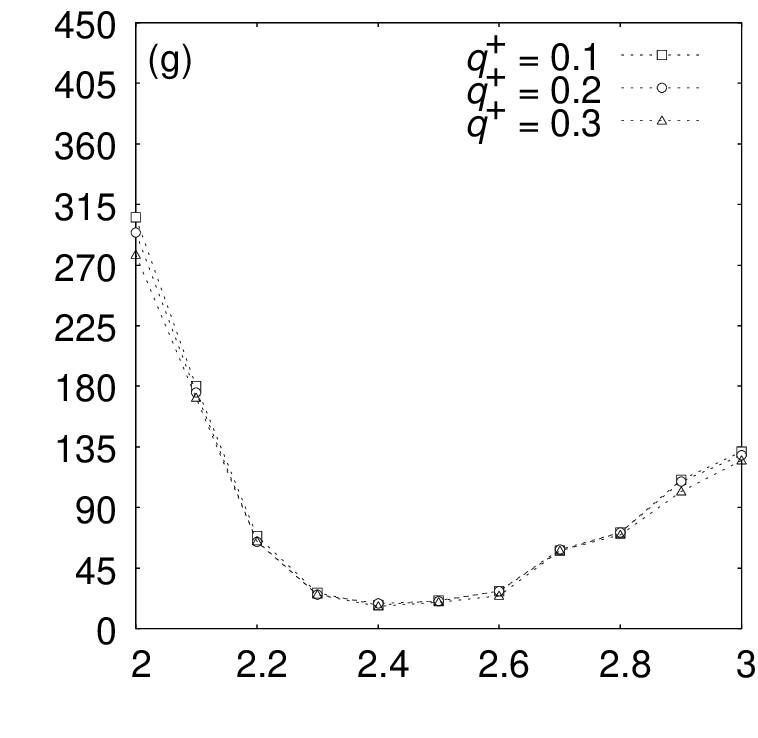} &
   \includegraphics[height=\graphHeight]{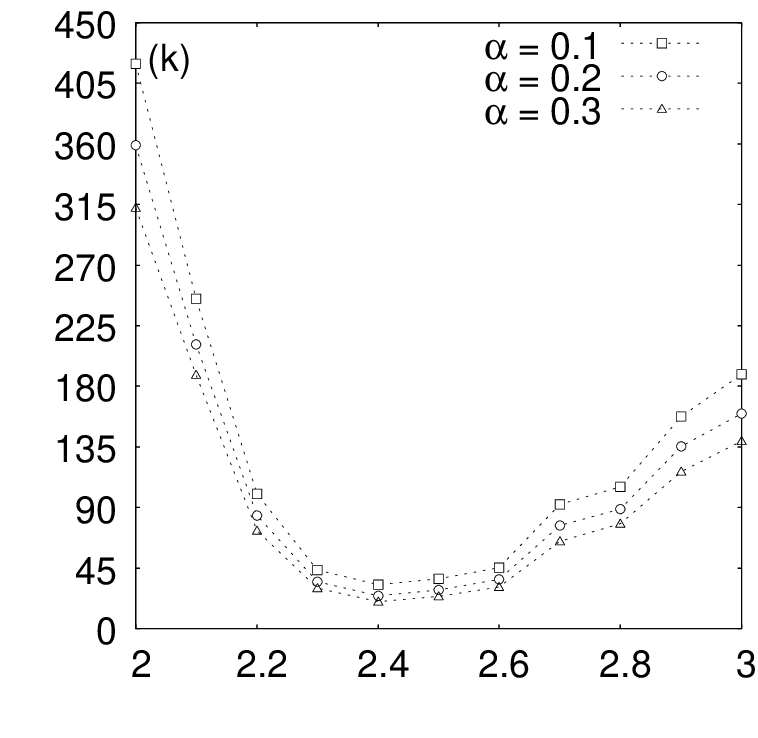} \\
   \includegraphics[height=\graphHeight]{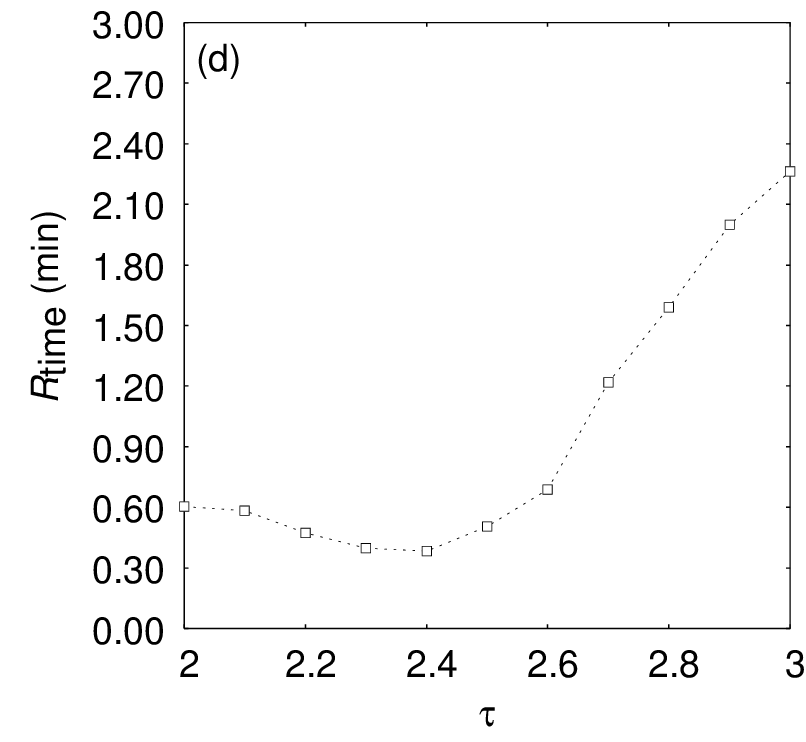} &
   \includegraphics[height=\graphHeight]{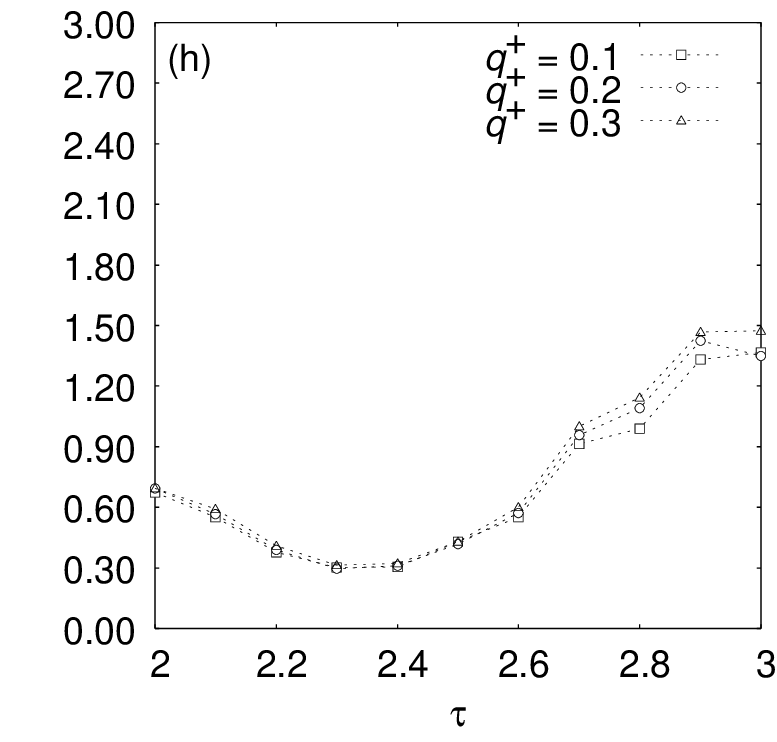} &
   \includegraphics[height=\graphHeight]{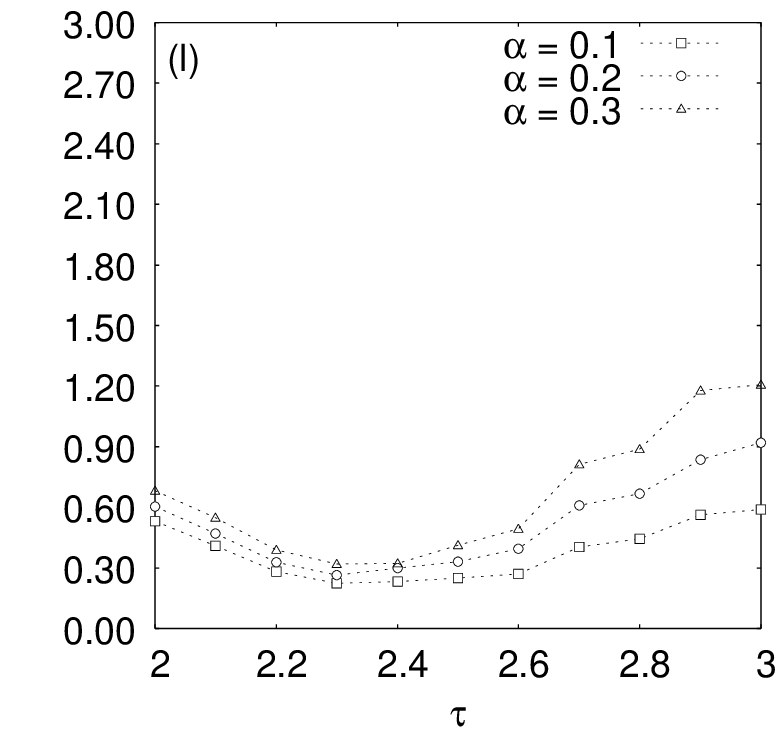}
   \end{tabular}
   \caption{Computational results for the GMZ heuristic (a--d), the VL heuristic
   (e--h), and the SB heuristic (i--l) when the halting condition is based on 
   the clustering coefficient. Plots refer to $\rconv$ (a, e, i), $\rswitch$ 
   (b, f, j), $\rw$ (c, g, k), and $\rtime$ (d, h, l).}
   \label{fig:mgmC}
\end{figure*}

Figure~\ref{fig:mgmC} shows the results obtained with this halting condition 
for the GMZ heuristic (parts~(a--d)), the VL heuristic (e--h), and the SB 
heuristic (i--l). The plots for $\rconv$ (Figure~\ref{fig:mgmC}(a, e, i)) show 
that $\rconv$ is close to $1$ for all the three heuristics, especially when 
$\tau \leq 2.1$ or $\tau \geq 2.8$. The plots for $\rswitch$ 
(Figure~\ref{fig:mgmC}(b, f, j)) show that the smallest value of $\rswitch$ is
obtained for $\tau \approx 2.4$, suggesting that the clustering coefficient 
converges faster for such a value of $\tau$. The parameters $q^+$ and $q^-$ of
the VL heuristic and $\alpha$ of the SB heuristic seem, curiously, to have small
impact on $\rswitch$. Furthermore, since $m$ is almost constant for $\tau \geq 
2.5$, $\rswitch$ does not seem to be proportional to $m$, as assumed in the 
analysis conducted in \cite{viger2005} for a slightly different power law.
The plots for $\rw$ (Figure~\ref{fig:mgmC}(c, g, k)) show that the smallest
value of $\rw$ is also obtained when $\tau \approx 2.4$, indicating that the
probability that an edge-switching attempt results in an unconnected $G$ is
smaller when $\tau \approx 2.4$. We note that the highest $\rw$ is obtained with
the SB heuristic. The reason for this behavior seems to be that both the GMZ
heuristic and the VL heuristic start with $w=1$, while the SB heuristic starts
with $w$ relatively close to $\rw$. The plots for $\rtime$
(Figure~\ref{fig:mgmC}(d, h, l)) show that the SB heuristic yields on average
the smallest running time, despite employing a more complex connectivity test.
For example, the SB heuristic has on average outperformed the GMZ heuristic by
roughly $12\%$ when $\tau = 2.0$, $44\%$ when $\tau = 2.3$, $61\%$ when
$\tau = 2.6$, and $74\%$ when $\tau=3.0$. In comparison to the VL heuristic,
these figures have been roughly $21\%$ when $\tau = 2.0$, $25\%$ when
$\tau = 2.3$, $51\%$ when $\tau = 2.6$, and $56\%$ when $\tau=3.0$. Regarding
the value of $\alpha$, the smallest average $\rtime$ for the SB heuristic
corresponds to $\alpha=0.1$. We expect $\rtime$ to decrease even more if we
continue decreasing $\alpha$, but this decrease will probably be progressively
smaller until an optimal value of $\alpha$ is achieved. Also, it is curious to
note that, for $\tau$ near $3.0$, the GMZ heuristic yields the smallest
$\rswitch$ but the highest $\rtime$ in comparison to the other heuristics. In
this situation, $\rw$ is so small that, even performing substantially fewer edge
switches, the ESMC method requires on average much longer to conclude.

\begin{figure*}[!t]
   \centering
   \begin{tabular}{rrr}
   \includegraphics[height=\graphHeight]{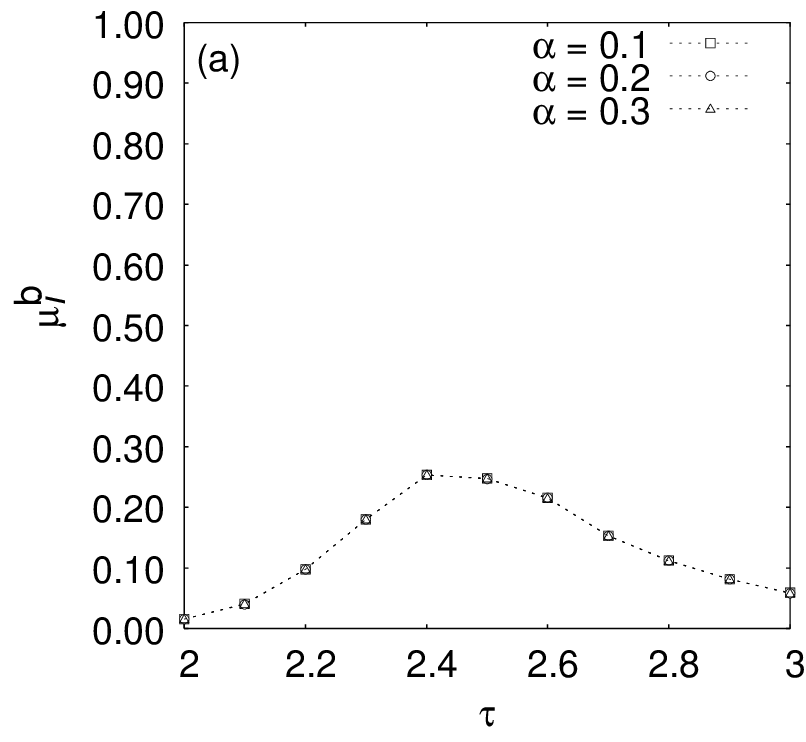} &
   \includegraphics[height=\graphHeight]{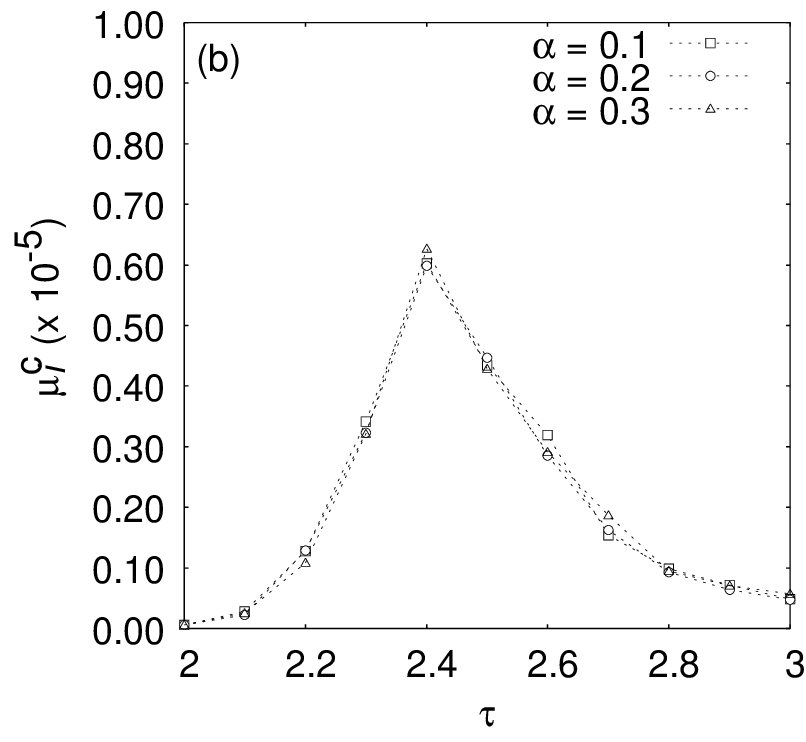} &
   \includegraphics[height=\graphHeight]{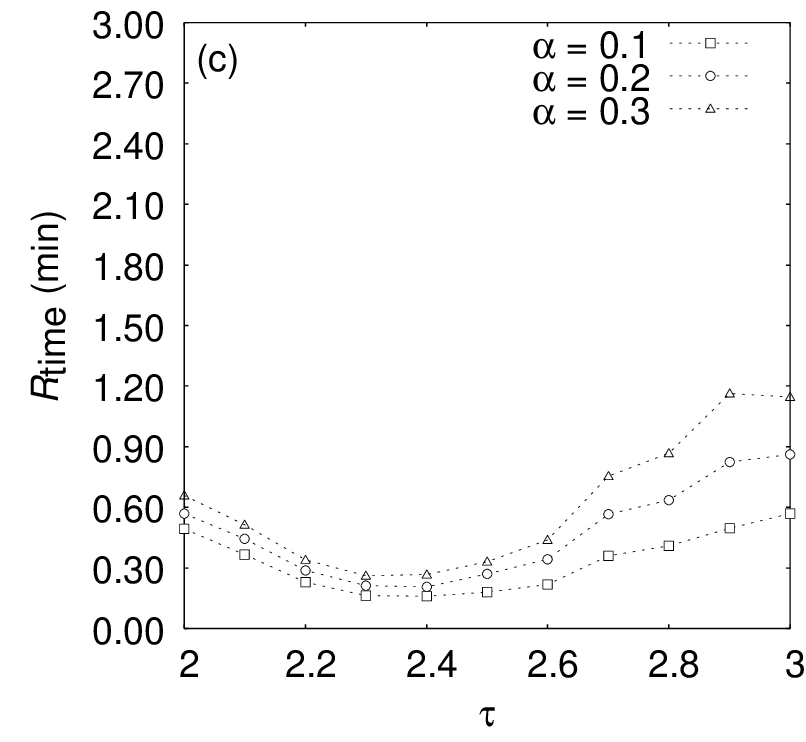}
   \end{tabular}
   \caption{Plots for $\pb_i$ (a), $\pc_i$ (b), and $\rtime$ when pair cuts 
           are ignored (c). The halting condition is the one based on the 
           clustering coefficient.}
   \label{fig:mgmCRho}
\end{figure*}

Figure~\ref{fig:mgmCRho}(a) presents the average $\pb_i$ at the end of an
execution for the SB heuristic when the halting condition is based on the
clustering coefficient. The value of $\tau$ for which we obtain the highest
average is $2.4$, in accordance with the fact that $\rw$ is on average minimum
for this same value (cf.\ Figure~\ref{fig:mgmC}(c, g, k)). When $\tau$ is
decreased from $2.4$, on average $\pb_i$ decreases as well, since the graph is
expected to have more edges and, consequently, fewer bridges. When $\tau$ is
increased from $2.4$, on average $\pb_i$ also decreases. The reason in this case
is that, since the number of edges remains practically constant as $\tau$ is
increased from $2.4$, and moreover the variance within the degree sequence
increases, the graph tends to acquire several star-like subgraphs and therefore
the fraction of adjacent or neighbor bridge pairs is expected to increase.
Figure~\ref{fig:mgmCRho}(b) refers to $\pc_i$. The behavior is similar, albeit
in an extremely smaller scale, thus indicating that the fraction of nonadjacent,
non-neighbor pair cuts in graphs whose node degrees are power-law-distributed is
on average negligible. If we ignore pair cuts and use $\rho_i=1-\pb_i$ in lieu
of (\ref{eq:rho}), then we obtain figures for $\rtime$ as shown in
Figure~\ref{fig:mgmCRho}(c). In this case $\rtime$ is on average significantly
smaller than when pair cuts are not ignored (Figure~\ref{fig:mgmC}(l)). This
decrease is on average higher when $d_1$ is expected to be smaller. Since for
small $d_1$ the time complexity of calculating the clustering coefficient is
relatively close to the time complexity of a connectivity test, speeding-up the
connectivity test impacts more strongly the overall running time. For example,
when $\tau=2.4$, in which case we have observed the value of $d_1$ to be
relatively small on average, ignoring pair cuts leads to a decrease in $\rtime$
of about $31\%$ on average. Likewise, when $\tau=2.0$, in which case we have
observed the opposite trend regarding the value of $d_1$, the decrease in
$\rtime$ is of about $7\%$.

%===============================================================================
\subsection{Halting on the average distance between nodes}

The second halting condition is based on letting $g(t)$ be the average distance 
between the nodes of $G$, which can be calculated by conducting a breath-first 
search rooted at each node of $G$. This calculation requires $\Theta(nm)$ 
time, therefore more than the calculation of the clustering coefficient. We have
used $\delta=30$ and $\gamma=10^{-3}$ for this halting condition. 

As we once again return to the issue raised at the end of Section~\ref{sec:newh}
on the convergence of $\bar \rho(t)$, for this second halting condition we
have observed the average value of $g(t)$ under the SB heuristic to stay below
roughly $1\%$ of the values obtained for the GMZ heuristic for all values of
$\tau$. As for the VL heuristic, the percentage remains the same but for
$\tau=2.1$ ($1.8\%$).

\begin{figure*}[!p]
   \centering
   \begin{tabular}{rrr}
   \includegraphics[height=\graphHeight]{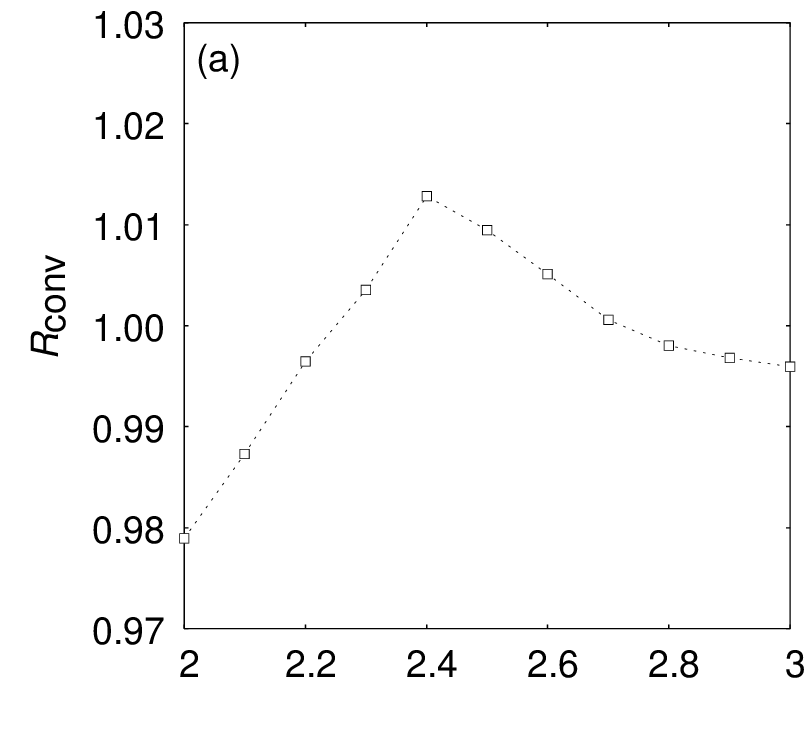} &
   \includegraphics[height=\graphHeight]{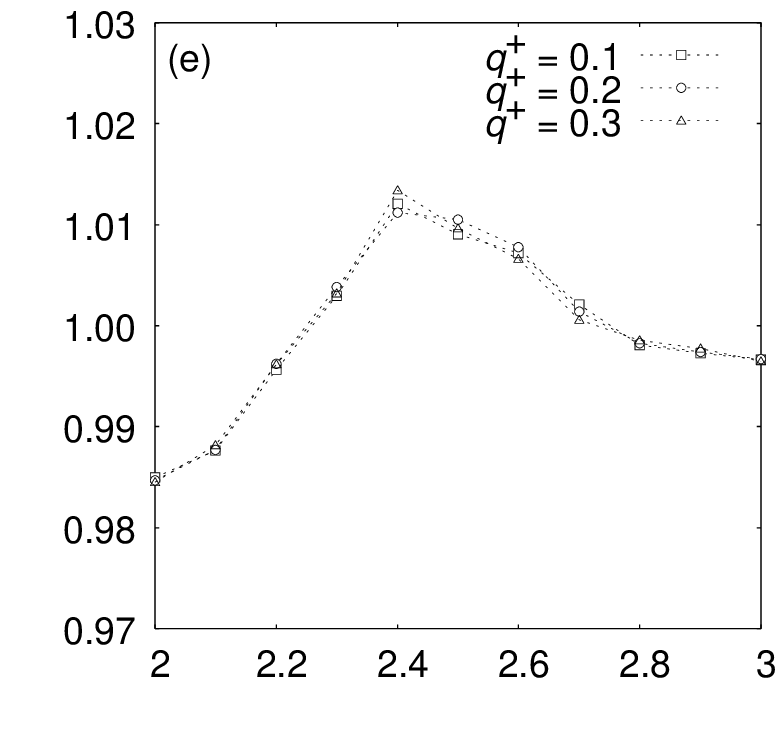} &
   \includegraphics[height=\graphHeight]{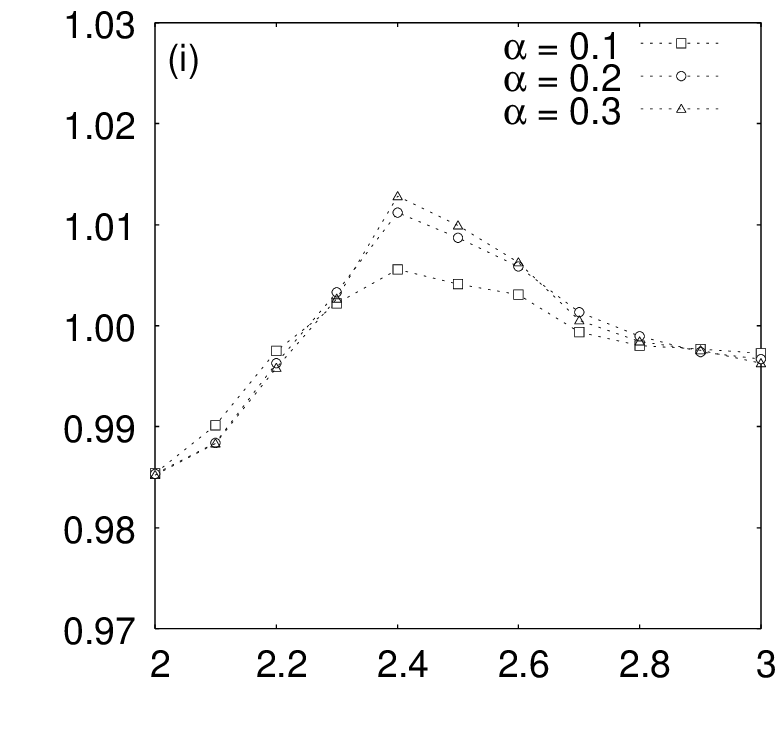} \\
   \includegraphics[height=\graphHeight]{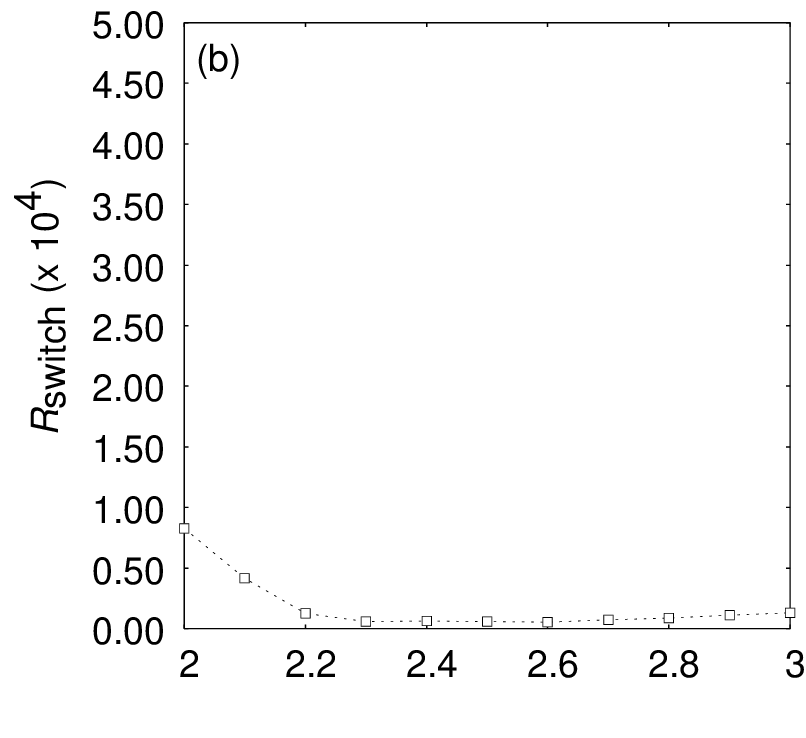} &
   \includegraphics[height=\graphHeight]{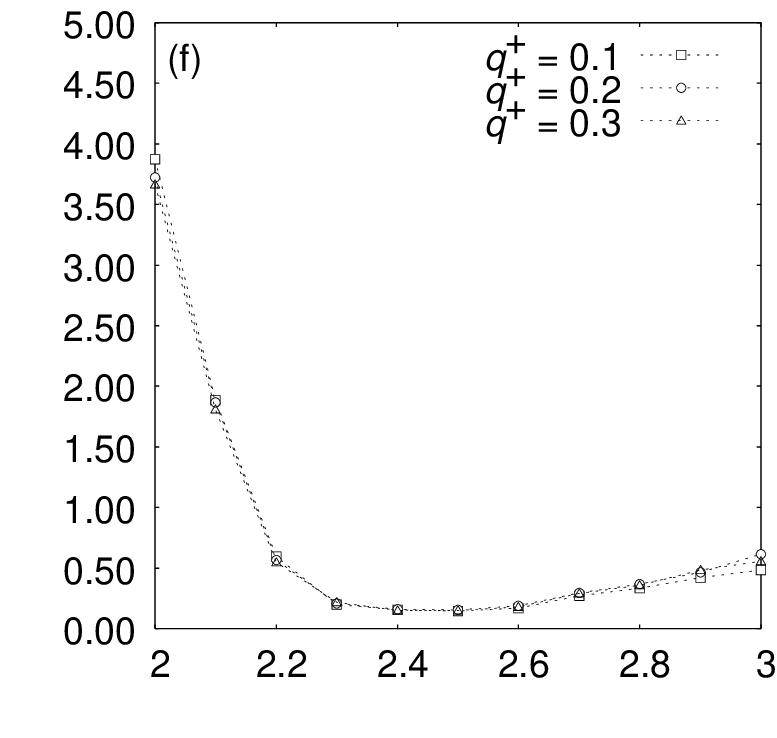} &
   \includegraphics[height=\graphHeight]{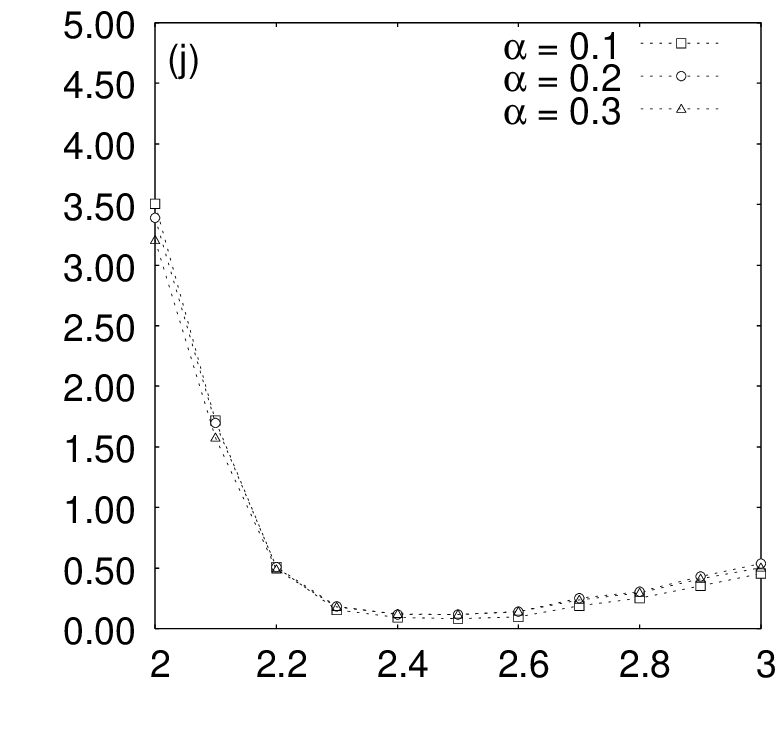} \\
   \includegraphics[height=\graphHeight]{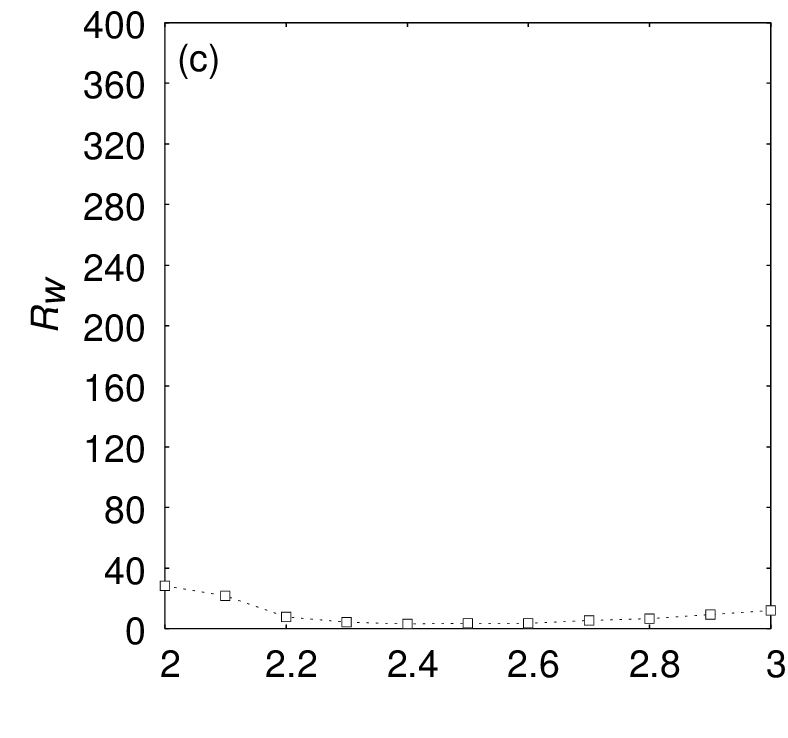} &
   \includegraphics[height=\graphHeight]{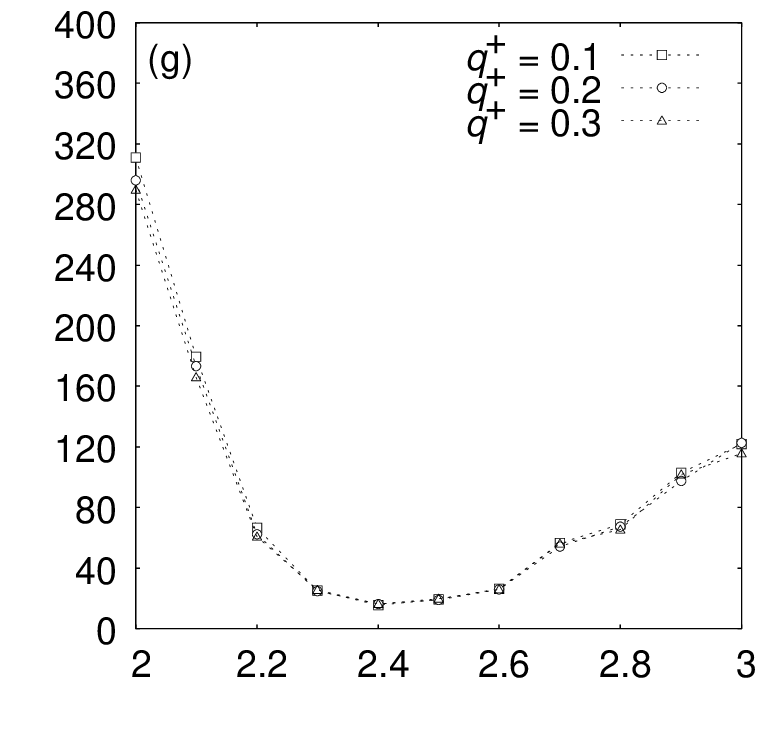} &
   \includegraphics[height=\graphHeight]{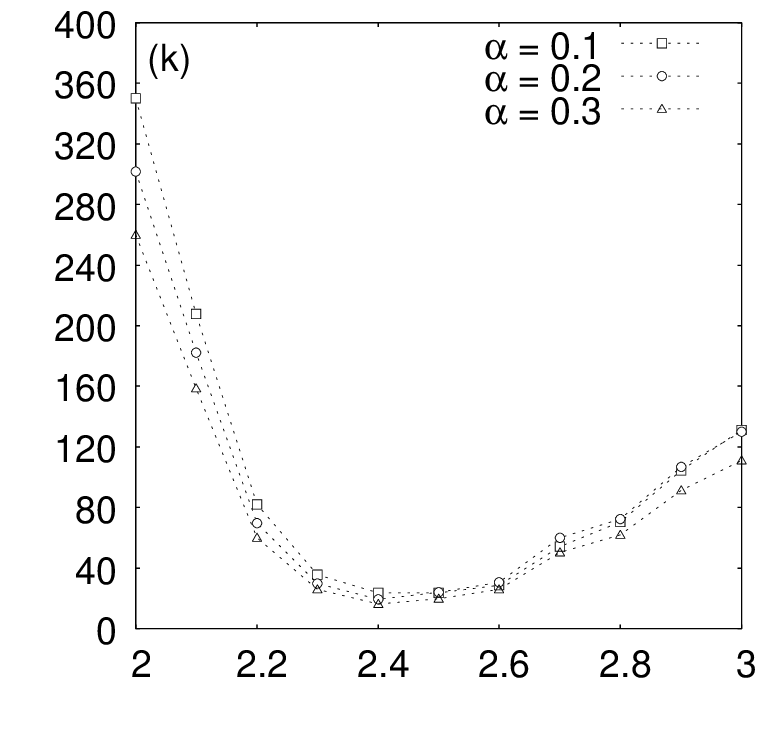} \\
   \includegraphics[height=\graphHeight]{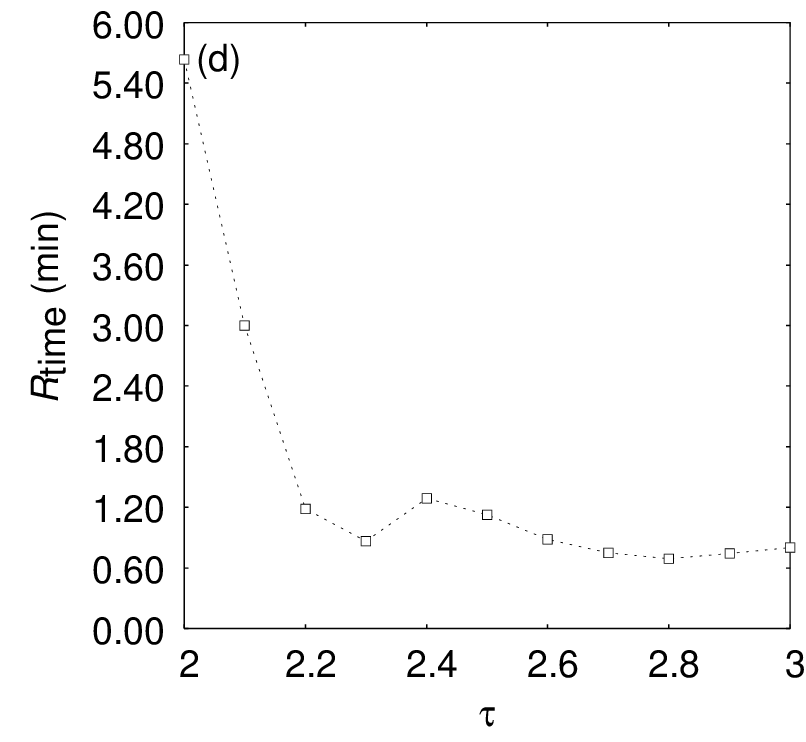} &
   \includegraphics[height=\graphHeight]{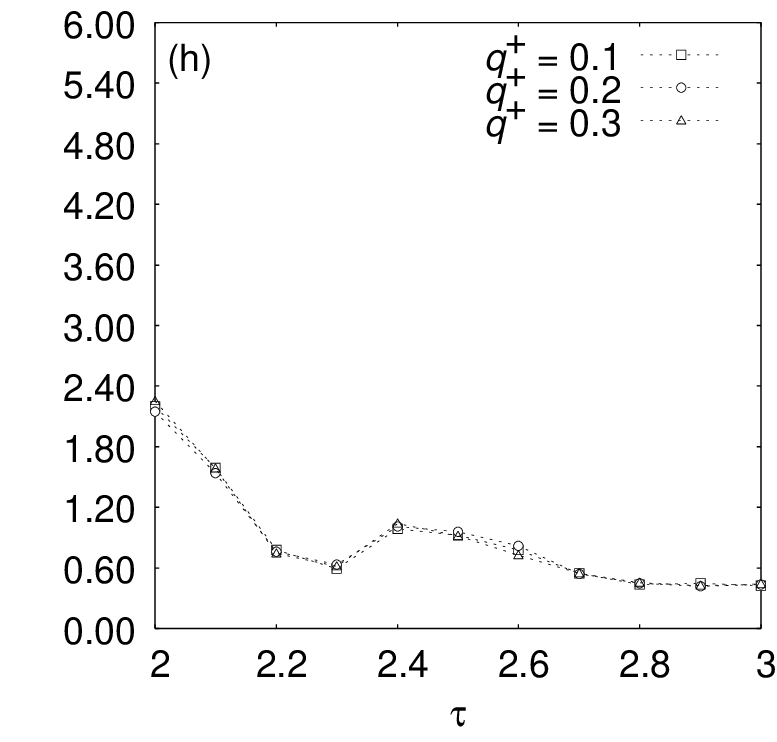} &
   \includegraphics[height=\graphHeight]{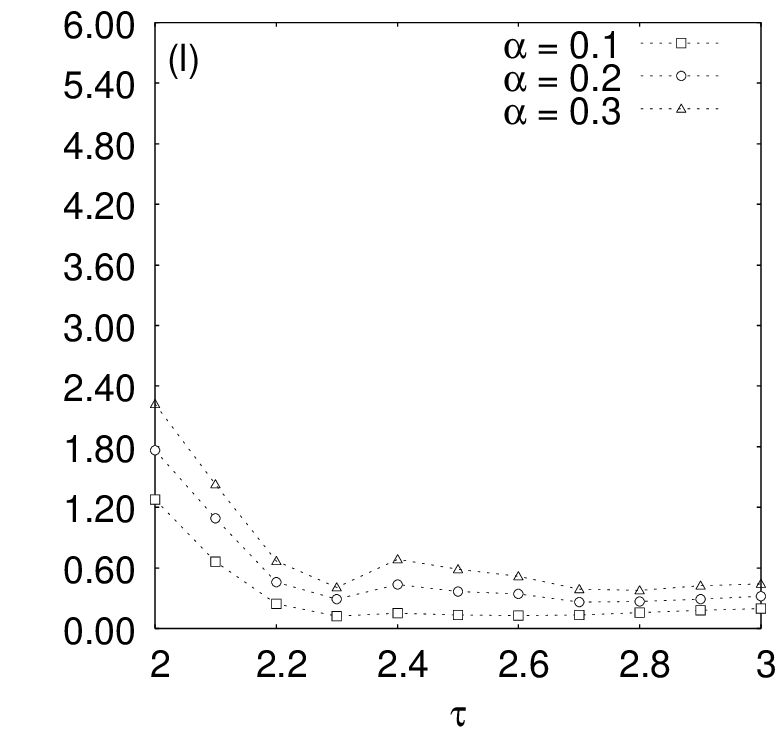}
   \end{tabular}
   \caption{Computational results for the GMZ heuristic (a--d), the VL heuristic
   (e--h), and the SB heuristic (i--l) when the halting condition is based on 
   the average distance between nodes. Plots refer to $\rconv$ (a, e, i),    
   $\rswitch$ (b, f, j), $\rw$ (c, g, k), and $\rtime$ (d, h, l).}
   \label{fig:mgmD}
\end{figure*}

Figure~\ref{fig:mgmD} shows the results when this is the halting condition for 
the GMZ heuristic (parts~(a--d)), the VL heuristic (e--h), and the SB heuristic
(i--l). The plots for $\rconv$ (Figure~\ref{fig:mgmD}(a, e, i)) show that
$\rconv$ is relatively far from $1$ in comparison to the results obtained with
the first halting condition (Figure~\ref{fig:mgmC}(a, e, i)). In order to obtain
$\rconv$ closer to $1$, we may need to increase $\delta$ and/or decrease
$\gamma$. Despite being not so close to $1$, the value of $\rconv$ is almost the
same regardless of which heuristic is used to adjust $w$. The plots for
$\rswitch$ (Figure~\ref{fig:mgmD}(b, f, j)) show that the smallest value of
$\rswitch$ occurs when $\tau \approx 2.4$. Similarly to the case of the
clustering coefficient, this suggests that the average distance between nodes
converges faster when $\tau \approx 2.4$.\footnote{This agreement of the three
heuristics under either halting condition may in fact be indicative that the
ESMC method itself converges faster for this value of $\tau$.} The plots for
$\rw$ (Figure~\ref{fig:mgmD}(c, g, k)) also show that the smallest $\rw$ is
obtained for $\tau \approx 2.4$. Regarding $\rtime$
(Figure~\ref{fig:mgmD}(d, h, l)), the plots show that the SB heuristic leads
once again to the smallest running time on average. For example, on average the
SB heuristic outperforms the GMZ heuristic by roughly $77\%$ when $\tau = 2.0$,
$86\%$ when $\tau = 2.3$, $85\%$ when $\tau = 2.6$, and $75\%$ when $\tau=3.0$.
In comparison to the VL heuristic, on average the SB heuristic outperforms it by
roughly $41\%$ when $\tau = 2.0$, $80\%$ when $\tau = 2.3$, $82\%$ when
$\tau = 2.6$, and $54\%$ when $\tau=3.0$. The average gain obtained with the SB
heuristic is higher under this halting condition, which can be explained by
noting that each transition is now slower than under the halting condition based
on the clustering coefficient. As a consequence, it is under the
average-distance halting condition that the impact of adjusting $w$ properly is
more strongly manifest. Also, and unlike what occurs with the first halting
condition, the gain obtained with the SB heuristic is now higher when $\tau$ is
around $2.4$. This suggests that the choice for $g(t)$ depends on a careful
consideration of each application's peculiarities. Regarding the value of
$\alpha$, the SB heuristic once again leads to the smallest $\rtime$ when
$\alpha=0.1$, suggesting that the optimal value of $\alpha$ is less than $0.1$.

\begin{figure*}[!t]
   \centering
   \begin{tabular}{rrr}
   \includegraphics[height=\graphHeight]{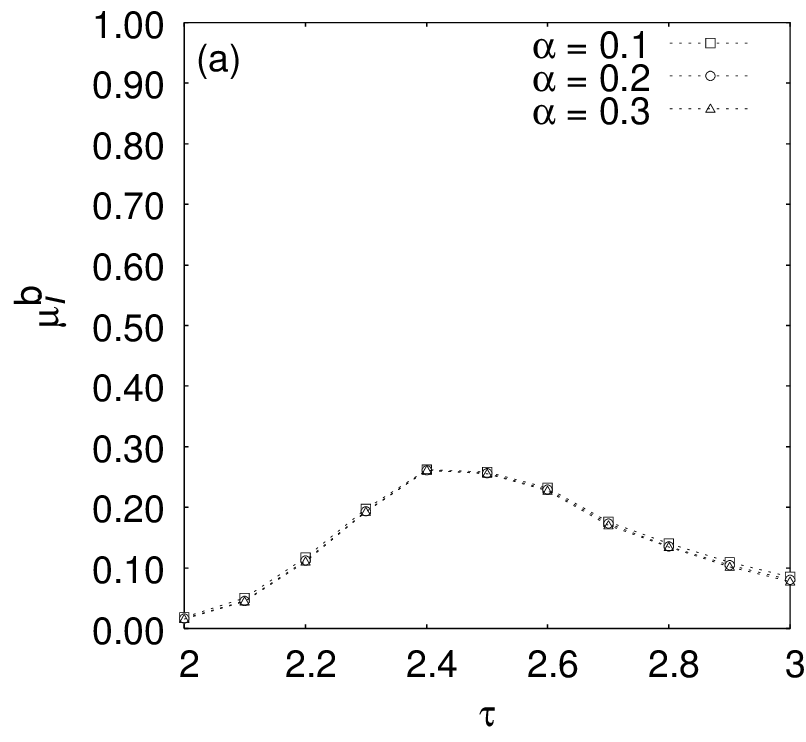} &
   \includegraphics[height=\graphHeight]{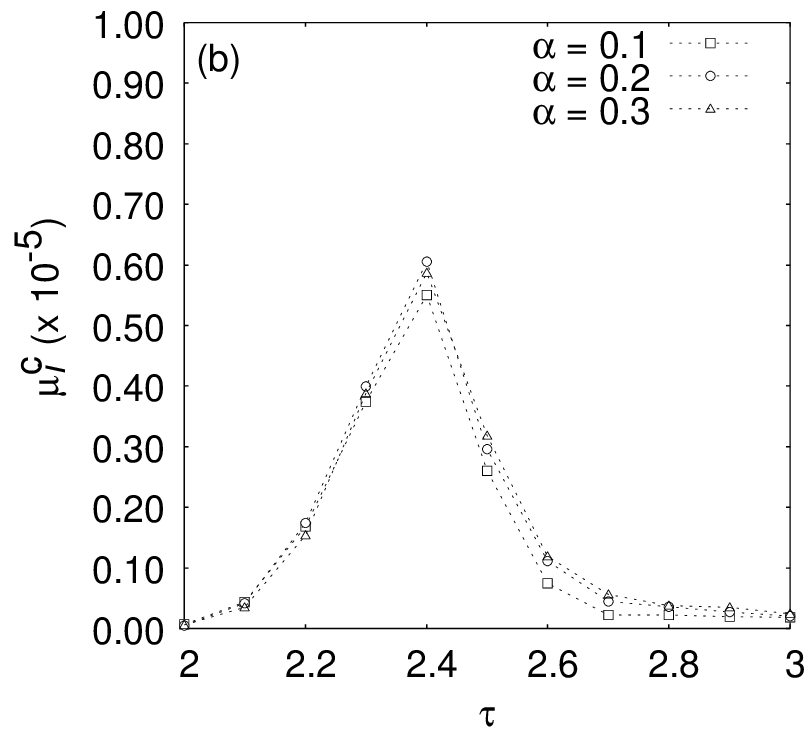} &
   \includegraphics[height=\graphHeight]{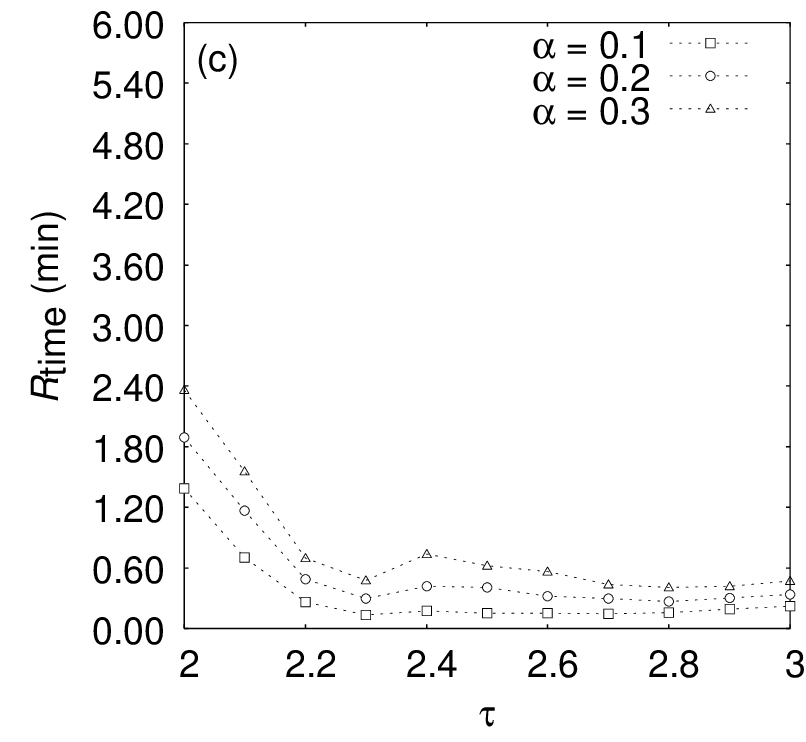}
   \end{tabular}
   \caption{Plots for $\pb_i$ (a), $\pc_i$ (b), and $\rtime$ when pair cuts 
           are ignored (c). The halting condition is the one based on the 
           average distance between nodes.}
   \label{fig:mgmDRho}
\end{figure*}

Figure~\ref{fig:mgmDRho} presents, respectively in parts (a) and (b), the
average $\pb_i$ and $\pc_i$ for the SB heuristic when the halting condition is
based on the average distance between nodes. The results are similar to the ones
shown in Figure~\ref{fig:mgmCRho}(a, b) for the halting condition based on the
clustering coefficient. The plots for $\rtime$ (Figure~\ref{fig:mgmDRho}(c)), on
the other hand, show a very different behavior. For almost all values of $\tau$,
$\rtime$ is now seen to increase slightly when pair cuts are ignored. The reason
for this behavior seems to be an insufficient number of samples. In fact, we
expect $\rtime$ to be very slightly smaller than that obtained when pair cuts
are not ignored. Since the time complexity of calculating the average distance
between nodes is significantly higher than that of a connectivity test, ignoring
pair cuts is therefore expected to have a small impact on the overall running
time of the method.

%===============================================================================
%===============================================================================
%===============================================================================
\section{Conclusions}  \label{sec:conclusion}

We have considered the problem of generating, uniformly at random, connected 
graphs that have a given degree sequence but no multiple edges or self-loops. 
We studied the ESMC method, which employs edge switches to transform a graph 
into another while preserving the degree sequence. This method consists of 
first deterministically  finding a graph with the desired properties and then 
performing random edge switches and also connectivity tests to obtain a 
randomized result.

We showed that, if we attempt to perform a constant number $w$ of edge switches 
between successive connectivity tests, then the method can be modeled as a 
Markov chain having a uniform stationary distribution. We also showed that, if 
$w$ is not constant but rather is adjusted as a function of the last 
connectivity test's outcome, then the method can still be modeled as a Markov 
chain of uniform stationary distribution. 

We have also introduced a new heuristic for adjusting $w$ that depends on the
probability that the graph being generated remains connected after an edge 
switch is attempted. In order to calculate this probability, we use a new
connectivity test that has the same time and space complexities as depth-first
search (cf.\ Appendix~\ref{sec:test}). Even though the resulting method cannot
always be modeled as a Markov chain, we showed that there are circumstances
under which it too converges to the uniform distribution.

One of the main issues regarding generation methods based on Markov chains 
is determining the number of transitions to be performed until the Markov chain 
is satisfactorily close to its stationary distribution. We have approached this 
issue by resorting to the pragmatic procedure of computing, after each 
transition, a certain function of the graph being generated, and halting the 
generation when the average of this function over all transitions seems to have 
converged. A proper choice for this function is essential to the efficacy of 
the method, but appears to require consideration on a case-by-case basis.

We have given computational results for power-law-based degree sequences. Our 
results contemplate two previous heuristics for adjusting $w$ and also our new 
heuristic, and were given for two distinct halting criteria. They show that our 
heuristic, on average, outperforms the two existing heuristics on power laws
for which $2 \leq \tau \leq 3$. 

The ESMC method can be especially useful to generate a group of connected random
graphs having the same degree sequence. After obtaining the first graph, we can 
continue performing a relatively small number of transitions to generate each 
additional instance, without having to run the method from its beginning.
Finally, the ESMC method can be extended to generate random graphs having a
given degree sequence and another desired property (e.g., graphs having the
clustering coefficient limited to a given interval). We need only find a means
of obtaining an initial graph having that property, then obtain an efficient
procedure to test whether a graph has that property, and also show the 
irreducibility of the Markov chain, that is, show that there is a sequence of 
edge-switching attempts connecting any two graphs having the given degree  
sequence and the desired property.

%===============================================================================
%===============================================================================
%===============================================================================
\subsection*{Acknowledgments}

The authors acknowledge partial support from CNPq, CAPES, and a FAPERJ BBP 
grant.

\bibliography{generation}
\bibliographystyle{plain}

\appendix
%===============================================================================
%===============================================================================
%===============================================================================
\section{The new connectivity test} \label{sec:test}

\begin{figure*}[!t]
   \centering
   \includegraphics[scale=\graphDraw]{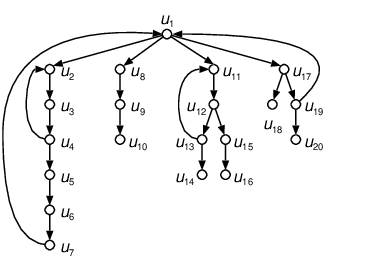}
   \caption{An example $S_i$. Nodes are visited in the order 
   $u_1,\ldots,u_{20}$.}
   \label{fig:conntest}
\end{figure*}

Given a graph $G_i$ of $\calg{D}$, we show how a modified depth-first search on
$G_i$ can be used to obtain $\rho_i$ in addition to testing whether $G_i$ is
connected. Let $S_i$ be the directed graph induced by a depth-first search on
$G_i$. This graph contains the same nodes as $G_i$ and a directed edge for each
edge of $G_i$. The direction of an edge in $S_i$ is the direction along which
the search traverses the edge for the first time. Let $(u_j \to u_k)$ be an edge
of $S_i$. We say that $u_j$ is the parent of $u_k$ (or, equivalently, $u_k$ is a
child of $u_j$) if the search visits $u_k$ for the first time from $u_j$. Edge
$(u_j \to u_k)$ is then called a tree edge, as it is part of a directed spanning
tree rooted at the start node of the search. If the search does not visit $u_k$
for the first time from $u_j$, then $(u_j \to u_k)$ is called a back edge, as it
necessarily represents a move toward an already visited node.
Figure~\ref{fig:conntest} shows an example $S_i$; nodes are numbered in such a
way that an edge is a tree edge if and only if it leads from a lower-numbered
node to a higher-numbered one.

The level of a node $u_j$ in $S_i$ is the length of the shortest directed path 
from the root to $u_j$. The descent and ancestry of $u_j$ in $S_i$ are, 
respectively, the set of nodes toward which a tree path exists from $u_j$ 
and the set of nodes from which a tree path exists toward $u_j$. Node $u_j$ 
is excluded from either set. Let $(u_j \to u_k)$ be a tree edge and $(u_x \to 
u_y)$ a back edge. We say that $(u_x \to u_y)$ covers $(u_j \to u_k)$ if 
$u_x=u_k$ or $u_x$ belongs to the descent of $u_k$, and furthermore $u_y=u_j$ 
or $u_y$ belongs to the ancestry of $u_j$. In Figure~\ref{fig:conntest}, edge 
$(u_4 \to u_2)$ covers edges $(u_2 \to u_3)$ and $(u_3 \to u_4)$. 

Let us proceed to the calculation of $\rho_i$, which by (\ref{eq:rho}) depends 
on $\pb_i$ and $\pc_i$.
 
%===============================================================================
\subsection{Handling bridge pairs}

Clearly, the number of nonadjacent, non-neighbor bridge pairs of $G_i$, on which
$\pb_i$ is based, can be obtained from the number of bridges of $G_i$, the
number of pairs of adjacent bridges of $G_i$, and the number of neighbor bridge
pairs of $G_i$. During the search, we count some undirected paths in $S_i$
(i.e., paths whose edges' directions are ignored) having certain special
properties. For each node $u_j$, we use the counters $\bridge{b}_j$,
$\bridge{bb}_j$, $\bridge{bbb}_j$, $\bridge{nb}_j$, and $\bridge{bnb}_j$ to
record how many undirected paths of $S_i$ start at $u_j$, proceed through nodes
in the descent of $u_j$ exclusively, and moreover consist in $G_i$ of,
respectively, one bridge, two bridges, three bridges, a non-bridge edge followed
by a bridge, and two bridges separated by a non-bridge edge. We now explain how
these counters can be used to obtain the number of nonadjacent, non-neighbor
bridge pairs of $G_i$ and also how we can calculate them during the search.

The number of bridges of $G_i$ can be easily obtained during the search, as an 
edge of $G_i$ is a bridge if and only if it is a tree edge of $S_i$ that is not 
covered by any back edge (e.g., $(u_1 \to u_8)$ in Figure~\ref{fig:conntest}). 
What we do is simply to accumulate $\bridge{b}_j$ into a global counter as the 
exploration of $u_j$ concludes. Obtaining the number of pairs of adjacent 
bridges of $G_i$ is also simple, since it is a matter of accumulating, as the 
exploration of $u_j$ concludes, the number of pairs of adjacent bridges that are
incident to $u_j$ and its descendants, that is,
\begin{equation}
   \binom{\bridge{b}_j}{2} + \bridge{bb}_j.
   \label{eq:abrid}
\end{equation}

As for obtaining the number of neighbor bridge pairs, note first that the edge 
connecting the two bridges can be of three types. It can be another bridge 
(e.g., $(u_1 \to u_8)$ connecting $(u_1 \to u_{11})$ to $(u_8 \to u_9)$, and 
$(u_8 \to u_9)$ connecting $(u_1 \to u_8)$ to $(u_9 \to u_{10})$ in 
Figure~\ref{fig:conntest}); it can be a tree edge that is not a bridge (e.g., 
$(u_1 \to u_{17})$ connecting $(u_1 \to u_{11})$ to $(u_{17} \to u_{18})$, and 
$(u_{11} \to u_{12})$ connecting $(u_1 \to u_{11})$ to $(u_{12} \to u_{15})$ in 
Figure~\ref{fig:conntest}); and, finally, it can be a back edge (e.g., 
$(u_{19} \to u_1)$ connecting $(u_1 \to u_{11})$ to $(u_{19} \to u_{20})$, and 
$(u_{13} \to u_{11})$ connecting $(u_1 \to u_{11})$ to $(u_{13} \to u_{14})$ in 
Figure~\ref{fig:conntest}). Let then $(u_j \to u_k)$ be a tree edge. As the 
exploration of $u_k$ concludes, for each $u_x$ in $u_k$'s descent from 
which a back edge exists toward $u_k$, we add $\bridge{b}_x$ to 
$\bridge{nb}_k$. We then accumulate
\begin{equation}
   \bridge{bb}_k(\bridge{b}_k-1) + \bridge{bbb}_k + \bridge{b}_k\bridge{nb}_k + \bridge{bnb}_k
   \label{eq:cbrid}
\end{equation} 
into the global counter of neighbor bridge pair of $G_i$. When at last the 
search returns to $u_k$'s parent $u_j$, we do one of the following: if $(u_j 
\to u_k)$ is a bridge, then we increment $\bridge{b}_j$ and add $\bridge{b}_k$ 
to $\bridge{bb}_j$, $\bridge{bb}_k$ to $\bridge{bbb}_j$, and $\bridge{nb}_k$ to 
$\bridge{bnb}_j$; otherwise, we add $\bridge{b}_k$ to $\bridge{nb}_j$.

%===============================================================================
\subsection{Handling pair cuts}

The number of nonadjacent, non-neighbor pair cuts, which is the basis for
computing $\pc_i$, can be obtained by calculating the number of pair cuts, the
number of adjacent pair cuts, and the number of neighbor pair cuts. Two edges of
$S_i$ form a pair cut if and only if they are covered by one single common back
edge (e.g., $(u_1 \to u_2)$ and $(u_4 \to u_5)$ in Figure~\ref{fig:conntest}).
In order to identify pair cuts during the search, for each node we store the
back edge that connects either the node itself or one of its descendants to its
lowest-level ancestor. If more than one back edge reaches the same node, then we
need store neither, since no edge through which the search is yet to backtrack
can be uniquely covered by any of them. 

Let $(u_j \to u_k)$ be a tree edge. Assume that $(u_j \to u_k)$ is covered only 
by the edge $(u_x \to u_y)$ and let $\paircut{x}{y}$ be a counter of the number 
of edges covered only by $(u_x \to u_y)$. Clearly, the number of pair cuts 
either covered by $(u_x \to u_y)$ or including this edge is
$\binom{\paircut{x}{y}+1}{2}$, since $(u_x \to u_y)$ also participates in a pair
cut along with each of the $\paircut{x}{y}$ edges that it covers. This number is
accumulated into a global counter of the pair cuts of $G_i$ as the search
detects that no edge through which it is yet to backtrack is covered only by
$(u_x \to u_y)$.

In order to identify adjacent and neighbor pair cuts, we need to keep some 
information regarding $(u_x \to u_y)$ and the edges covered only by it as the 
search backtracks from $u_k$. Besides $(u_x \to u_y)$ itself and 
$\paircut{x}{y}$, we also need to retain information on three other nodes, which
we denote by $v_1$, $v_2$, and $v_3$. Nodes $v_1$ and $v_2$ are the two
lowest-level nodes such that the edge between each of them and its parent is
covered only by $(u_x \to u_y)$. Node $v_3$ is the highest-level node such that
the edge between it and its parent is covered only by $(u_x \to u_y)$. For
example, as the search backtracks from $u_3$ in the case of
Figure~\ref{fig:conntest}, we store the back edge $(u_7 \to u_1)$ and let
$v_1=u_5$, $v_2=u_6$, and $v_3=u_7$.

\begin{figure*}[!t]
   \centering
   \begin{tabular}{lllll}
   \includegraphics[scale=\graphDraw]{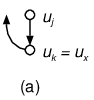} &
   \includegraphics[scale=\graphDraw]{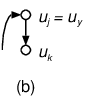} &
   \includegraphics[scale=\graphDraw]{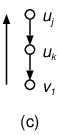} &
   \includegraphics[scale=\graphDraw]{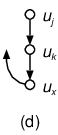} &
   \includegraphics[scale=\graphDraw]{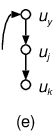} \\
   \includegraphics[scale=\graphDraw]{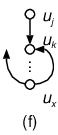} &
   \includegraphics[scale=\graphDraw]{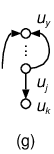} &
   \includegraphics[scale=\graphDraw]{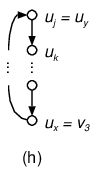} &
   \includegraphics[scale=\graphDraw]{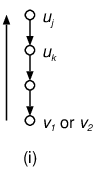} &
   \includegraphics[scale=\graphDraw]{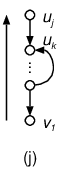} 
   \end{tabular}
   \caption{Scenarios for the occurrence of adjacent and neighbor pair cuts.}
   \label{fig:pcex}
\end{figure*}

Assume now that the search has concluded the exploration of all the neighbors of
$u_j$. In the case of a child $u_k$ of $u_j$, assume as above that
$(u_j \to u_k)$ is covered only by the back edge $(u_x \to u_y)$. Adjacent pair
cuts can be identified in three scenarios: when $u_k=u_x$
(Figure~\ref{fig:pcex}(a)), when $u_j=u_y$ (Figure~\ref{fig:pcex}(b)), and when
$u_k$ is the parent of $v_1$ (Figure~\ref{fig:pcex}(c)). As for neighbor pair
cuts, there are five cases. The first case happens when a tree edge connects
$(u_x \to u_y)$ to one of the tree edges covered only by it; this can be
identified either when $u_x$ is a child of $u_k$ (Figure~\ref{fig:pcex}(d)) or
when $u_y$ is the parent of $u_j$ (Figure~\ref{fig:pcex}(e)). The second case
occurs when another back edge connects $(u_x \to u_y)$ to one of the tree edges
covered only by it; this can be identified either by the existence of the back
edge $(u_x \to u_k)$ (Figure~\ref{fig:pcex}(f)) or by the existence of the back
edge $(u_j \to u_y)$ (Figure~\ref{fig:pcex}(g)). The third case occurs when
$(u_x \to u_y)$ connects two edges covered only by it, and can be identified
when $u_j=u_y$ and $u_x=v_3$ (Figure~\ref{fig:pcex}(h)). The fourth case occurs
when a tree edge connects two other tree edges, the latter two covered only by
$(u_x \to u_y)$; this case can be identified by $v_1$ or $v_2$ being two levels
above $u_k$ (Figure~\ref{fig:pcex}(i)). The fifth and last case happens when
another back edge connects two edges covered only by $(u_x \to u_y)$, which can
be identified by the existence of a back edge from the parent of $v_1$ to $u_k$ 
(Figure~\ref{fig:pcex}(j)). After updating the number of adjacent pair cuts and 
neighbor pair cuts before the search backtracks from $u_j$, we increment 
$\paircut{x}{y}$ and let $v_2=v_1$ and $v_1=u_k$. If no node is currently 
marked as $v_3$, then we also let $v_3=u_k$.

%===============================================================================
\subsection{Complexity}

Let us now discuss the space and time complexities of this modified depth-first 
search.  Clearly, the ESMC method requires $\Omega(m)$ space, since we need to 
store an array with the edges of the graph being generated. During the search, 
for each node $u_k$, we need to store its parent (say, $u_j$), its level, the 
$\bridge{}_k$'s, and the back edge covering $(u_j \to u_k)$ that reaches the 
lowest-level node. If $(u_x \to u_y)$ is this edge, then we also need to store
the $v_1$, $v_2$, $v_3$, and $\paircut{x}{y}$ corresponding to $(u_j \to u_k)$. 
Summing up over all nodes, this information requires only $\Theta(n)$ space. 
Furthermore, for each node we keep a list of the back edges arriving at it for 
the sake of handling the cases in Figure~\ref{fig:pcex}(f, j), which requires 
$O(m)$ space overall. We also, finally, keep a global $n$-element array for
nodes to register the back edges originating at them. This is needed for
identifying the occurrence of the scenario illustrated in
Figure~\ref{fig:pcex}(g). We then see, in summary, that the modified depth-first
search does not change the space complexity of the ESMC method. 

Obtaining the time complexity requires that we detail the steps performed 
during the exploration of a node $u_j$ of $G_i$. First we explore each neighbor
$u_k$ of $u_j$, and update the $\bridge{}_j$'s if $(u_j \to u_k)$ is a tree 
edge. Otherwise, if $(u_j \to u_k)$ is a back edge, then we include it in the 
list of back edges arriving at $u_k$ and record the back edge that leaves $u_j$
and arrives at its lowest-level ancestor. After exploring the entire descent of 
$u_j$, for each node toward which there is a back edge leaving $u_j$ we set a 
mark in the $n$-element array. Then, for each child $u_k$ of $u_j$, we update
the counters of pair cuts using the $n$-element array, the information regarding
the back edge, say $(u_x \to u_y)$, that reaches the lowest-level node, and the
list of back edges arriving at $u_k$. The edge $(u_x \to u_y)$ may become the
back edge that arrives at $u_j$'s lowest-level ancestor; in this case,
we also update $\paircut{x}{y}$, $v_1$, $v_2$, and $v_3$, 
which requires $O(1)$ time. We then reset all marks in the $n$-element 
array,\footnote{Note that at this moment only neighbors of $u_j$ may be marked. 
So this step can be performed without checking all $n$ positions.} and for each 
back edge arriving at $u_j$ we update $\bridge{nb}_j$, which also requires only 
$O(1)$ time for each edge. Finally, we conclude the exploration of $u_j$ by 
updating the counters of adjacent and neighbor bridge pairs using 
(\ref{eq:abrid}) and (\ref{eq:cbrid}). We then see that a tree edge $(u_j \to 
u_k)$ is visited at most three times, twice when $u_j$ and $u_k$ are exploring 
their neighbors, and once more when $u_j$ revisits the tree edges leaving it to 
update the pair-cut counters and the back edge reaching the lowest-level node.
Each back edge $(u_x \to u_y)$, in turn, is visited at most six times, twice
when $u_x$ and $u_y$ are exploring their neighbors, twice when $u_x$ set and
reset marks in the $n$-element array, once when updating $\bridge{nb}_y$, and
once more when the parent of $u_y$ is updating the number of neighbor pair cuts 
(cf.\ the cases illustrated in Figure~\ref{fig:pcex}(f, j)). In conclusion, the 
time complexity of the modified  depth-first search is $O(m)$, thus the same 
as that of the standard depth-first search.

\end{document}